\documentclass[aps,prd,floatfix,superscriptaddress,preprintnumbers,nofootinbib,preprint,tightenlines,longbibliography]{revtex4-1}
\pdfoutput=1
\usepackage[english]{babel}
\usepackage{amsmath,amssymb,amsbsy,amstext,amsthm,booktabs,exscale,relsize,slashed,graphicx,amsfonts,upgreek,xcolor}
\usepackage{multirow,dcolumn,bm,enumerate,url,hyperref}

\hypersetup{
  colorlinks   = true, 
  urlcolor     = magenta, 
  linkcolor    = magenta, 
  citecolor   = magenta 
}

\newcommand{\Mstar}{M_\star}
\newcommand{\Rstar}{R_\star}
\newcommand{\mdm}{m_{\rm DM}}

\newcommand{\rhoDM}{\rho_{\rm DM}}

\newcommand{\sigvee}{\langle \sigma_{\rm ann} v \rangle}

\newcommand{\gev}{{\rm GeV}}
\newcommand{\mev}{{\rm MeV}}

\definecolor{lightsabergreen}{rgb}{.3,.8,.5}

\newcommand{\acro}[1]{\textsc{\MakeLowercase{#1}}}

\begin{document}

\title{Warming Nuclear Pasta with Dark Matter: \\
Kinetic and Annihilation Heating of Neutron Star Crusts}
\preprint{MIT-CTP/5152}

\author{Javier F. Acevedo}
\thanks{{\scriptsize Email}: \href{mailto:17jfa1@queensu.ca}{17jfa1@queensu.ca}; {\scriptsize ORCID}: \href{http://orcid.org/0000-0003-3666-0951}{0000-0003-3666-0951}}
\affiliation{\smaller The Arthur B. McDonald Canadian Astroparticle Physics Research Institute, \protect\\ Department of Physics, Engineering Physics, and Astronomy, \protect\\ Queen's University, Kingston, Ontario, K7L 2S8, Canada}

\author{Joseph Bramante}
\thanks{{\scriptsize Email}: \href{mailto:jab23@queensu.ca}{jab23@queensu.ca}; {\scriptsize ORCID}: \href{http://orcid.org/0000-0001-8905-1960}{0000-0001-8905-1960}}
\affiliation{\smaller The Arthur B. McDonald Canadian Astroparticle Physics Research Institute, \protect\\ Department of Physics, Engineering Physics, and Astronomy, \protect\\ Queen's University, Kingston, Ontario, K7L 2S8, Canada}
\affiliation{\smaller Perimeter Institute for Theoretical Physics, Waterloo, ON N2J 2W9, Canada}

\author{Rebecca K. Leane}
\thanks{{\scriptsize Email}: \href{mailto:rleane@mit.edu}{rleane@mit.edu}; {\scriptsize ORCID}: \href{http://orcid.org/0000-0002-1287-8780}{0000-0002-1287-8780}}
\affiliation{\smaller Center for Theoretical Physics, Massachusetts Institute of Technology, Cambridge, MA 02139, USA}

\author{Nirmal Raj}
\thanks{{\scriptsize Email}: \href{mailto:nraj@triumf.ca}{nraj@triumf.ca}; {\scriptsize ORCID}: \href{http://orcid.org/0000-0002-4378-1201}{0000-0002-4378-1201}}
\affiliation{\smaller TRIUMF, 4004 Wesbrook Mall, Vancouver, BC V6T 2A3, Canada}

\begin{abstract}

Neutron stars serve as excellent next-generation thermal detectors of dark matter, heated by the scattering and annihilation of dark matter accelerated to relativistic speeds in their deep gravitational wells. 
However, the dynamics of neutron star cores are uncertain, making it difficult at present to unequivocally compute dark matter scattering in this region. On the other hand, the physics of an outer layer of the neutron star, the crust, is more robustly understood.  We show that dark matter scattering solely with the low-density crust still kinetically heats neutron stars to infrared temperatures detectable by forthcoming telescopes. 
We find that for both spin-independent and spin-dependent scattering on nucleons, the crust-only cross section sensitivity is $10^{-43} - 10^{-41}$~cm$^2$ for dark matter masses of 100 MeV $-$ 1 PeV, with 
the best sensitivity arising from dark matter scattering with a crust constituent called nuclear pasta (including gnocchi, spaghetti, and lasagna phases). For dark matter masses from 10 eV to 1 MeV, the sensitivity is $10^{-39} - 10^{-34}$~cm$^2$, arising from exciting collective phonon modes in a neutron superfluid in the inner crust.
Furthermore, for any $s$-wave or $p$-wave annihilating dark matter, we show that dark matter will efficiently annihilate by thermalizing just with the neutron star crust, regardless of whether the dark matter ever scatters with the neutron star core. This implies efficient annihilation in neutron stars for any electroweakly interacting dark matter with inelastic mass splittings of up to 200 MeV, including Higgsinos. 
We conclude that neutron star crusts play a key role in dark matter scattering and annihilation in neutron stars.
\end{abstract}

\maketitle

\tableofcontents

\newpage 

\section{Introduction}

The existence of dark matter is well established~\cite{Bertone:2004pz}. 
However, despite extensive experimental efforts to unveil dark matter's microscopic properties, its identity remains elusive.
One promising approach to unmask non-gravitational dark matter interactions is to observe their impact on the ultradense interiors of neutron stars and white dwarfs~\cite{
Goldman:1989nd,
Gould:1989gw,
Kouvaris:2007ay,
Bertone:2007ae,
deLavallaz:2010wp,
Kouvaris:2010vv,
McDermott:2011jp,
Kouvaris:2011fi,
Guver:2012ba,
Bramante:2013hn,
Bell:2013xk,
Bramante:2013nma,
Bertoni:2013bsa,
Kouvaris:2010jy,
McCullough:2010ai,
Perez-Garcia:2014dra,
Bramante:2015cua,
Graham:2015apa,
Cermeno:2016olb,
Graham:2018efk,
Acevedo:2019gre,
Janish:2019nkk,
Krall:2017xij,
McKeen:2018xwc}. 
In particular, it has recently been suggested~\cite{Baryakhtar:2017dbj,
Raj:2017wrv,
Bell:2018pkk,
Garani:2018kkd,
Chen:2018ohx,
Garani:2018kkd,
Hamaguchi:2019oev,
Camargo:2019wou,
Bell:2019pyc,
Garani:2019fpa,
Joglekar:2019vzy} 
that old neutron stars can be heated to infrared temperatures via gravitationally accelerated dark matter depositing their kinetic energy.
This signature is largely model-independent, and potentially detectable with forthcoming infrared telescopes, such as the James Webb Space Telescope, the Thirty Meter Telescope, and the  Extremely Large Telescope \cite{Baryakhtar:2017dbj}. This requires that a suitable neutron star candidate at a distance $\lesssim$ 100 pc from Earth is found; pulsars have already been found at distances of $\sim 100~{\rm pc}$~\cite{Manchester:2004bp}. 
Measuring an old neutron star's infrared luminosity would provide unprecedented sensitivity to dark matter interactions with nucleons and leptons.

Previous work on neutron star capture of dark matter had assumed that dark matter is captured by the ultradense neutron star core of radius $\simeq$ 10 km, and consisting of neutrons, protons, electrons, and muons.
These treatments neglect the scattering of dark matter with the stellar crust, which is on average orders of magnitude less dense than the core and is only $\sim$ 1 km thick. The justification for this approximation is that the optical depth of the core is far greater than that of the crust.

In this work, we will revisit this approximation, and in particular consider dark matter capture {\em exclusively} through scattering with crust constituents.
Investigating neutron star crust capture is important for several reasons:
\begin{enumerate}

\item The high-density phase of matter in a neutron star core is currently uncertain~\cite{Page:2006ud}. 
It has been hypothesized that the core could consist of 
meson and hyperon condensates ~\cite{1975ApJ...199..471H,1982ApJ...258..306H,BROWN19761,Page:2006ud},
or deconfined $ud$ quark matter~\cite{Holdom:2017gdc} 
or $uds$ quark matter~\cite{Ivanenko1965,
Ivanenko1969,
Collins:1974ky,
Weber:1999qj,
Lastowiecki2015,
Chatterjee:2015pua}; 
the latter could exist in a color flavor locked phase~\cite{Alford:1998mk,Alford:2007xm}
and may co-exist with
confined states~\cite{Glendenning:1995rd,Bedaque:2001je,Kaplan:2001qk}.  
While the existence of any additional degrees of freedom in the neutron star generally lead to softer equations of state, and therefore predict a reduced neutron star mass, the heaviest found neutron stars to date, i.e., PSR J0740+6620 at $2.14^{+0.10}_{-0.09}\, M_\odot$ \cite{Cromartie:2019kug}, still permit exotic cores~\cite{RikovskaStone:2006ta,Chatterjee:2015pua,Haensel:2016pjp,Motta:2019ywl}.
Dark matter scattering on these phases may be kinematically forbidden~\cite{Bertoni:2013bsa}. 

Moreover, as the sensitivity to dark matter parameters will be ultimately obtained by observing individual neutron stars, it would be very difficult (if not impossible) to ascertain whether or not a particular star's core had undergone a phase transition to exotic matter.
 
\item There is no empirical knowledge of the inner structure of neutron stars, and although there is consensus among theoretical models, this is subject to change as the astrophysics of compact stars is a rapidly evolving field.
Therefore, it is crucial to estimate the sensitivity to dark matter detection for every layer of the neutron star interior using current models.
This sensitivity, as determined by scattering on all neutron star layers, will be more robust against future evolution of nuclear/astrophysical theories pertaining to any subset of these layers.

\item In some scenarios, dark matter quickly attains thermal equilibrium with the neutron star, and settles into a small volume in the core. However, dark matter with scattering cross sections suppressed at low momentum will remain oscillating at semi-relativistic speeds within the entire volume of the star. 
Treating this scenario requires detailed understanding of dark matter scattering with the neutron star crust. 
For this ``partial thermalization'' scenario, we will find that dark matter can still annihilate efficiently and heat the neutron star up to detectable infrared temperatures higher than those induced by kinetic/capture heating alone. 
This feature of neutron star crust capture significantly raises the discovery prospects of dark matter models with large inter-state mass splittings, {\em i.e.} inelastic dark matter models, such as the supersymmetric Higgsino.

\item Dark matter itself could be a particle that is preferentially captured in the neutron star crust, {\em e.g.} if its scattering is sensitive to the environmental density~\cite{Boddy:2012xs}, or if it is strongly interacting so that it is guaranteed to scatter with the crust before reaching the core during its transit through the star~\cite{Baryakhtar:2017dbj}. 

\end{enumerate}
%

This paper is structured as follows. 
In Section~\ref{sec:struc}, we first provide a layer-by-layer overview of neutron star structure, including the crust.
In Section~\ref{sec:revcap}, we provide a general review of dark matter capture and kinetic heating in neutron stars.  
In Section~\ref{sec:capture}, we investigate in detail dark matter capture in the crust, describing the various scattering processes by which dark matter deposits energy while transiting the crust; we also estimate the scattering cross section sensitivities of crustal capture and compare them with constraints from underground dark matter experiments.
In Section~\ref{sec:ann}, we investigate the impact of heating via dark matter annihilations. 
Our work on dark matter-pasta interactions boils down to some important conclusions, which we sift through in Section~\ref{sec:conc}.
Detailed descriptions of the neutron star crust, stellar temperatures due to dark matter kinetic heating, and dark matter thermalization with the crust are collected in the appendix.

\begin{figure}[t] 
\includegraphics[scale=0.895]{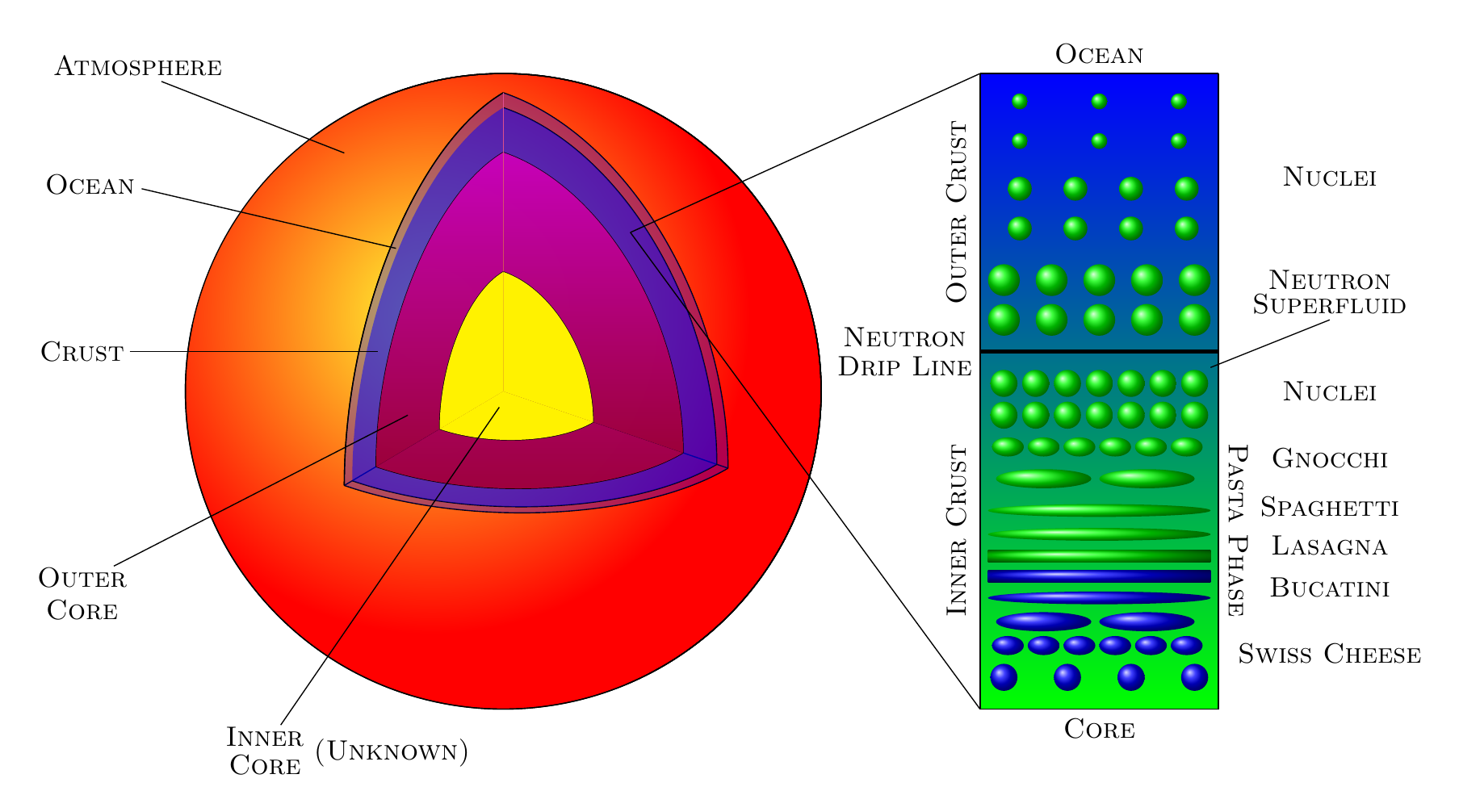} 
\caption{Interior structure of a neutron star, with a zoom-in on the crust (not to scale).}
\label{fig:nsinterior1}
\end{figure}

\begin{figure} 
\includegraphics[scale=0.66]{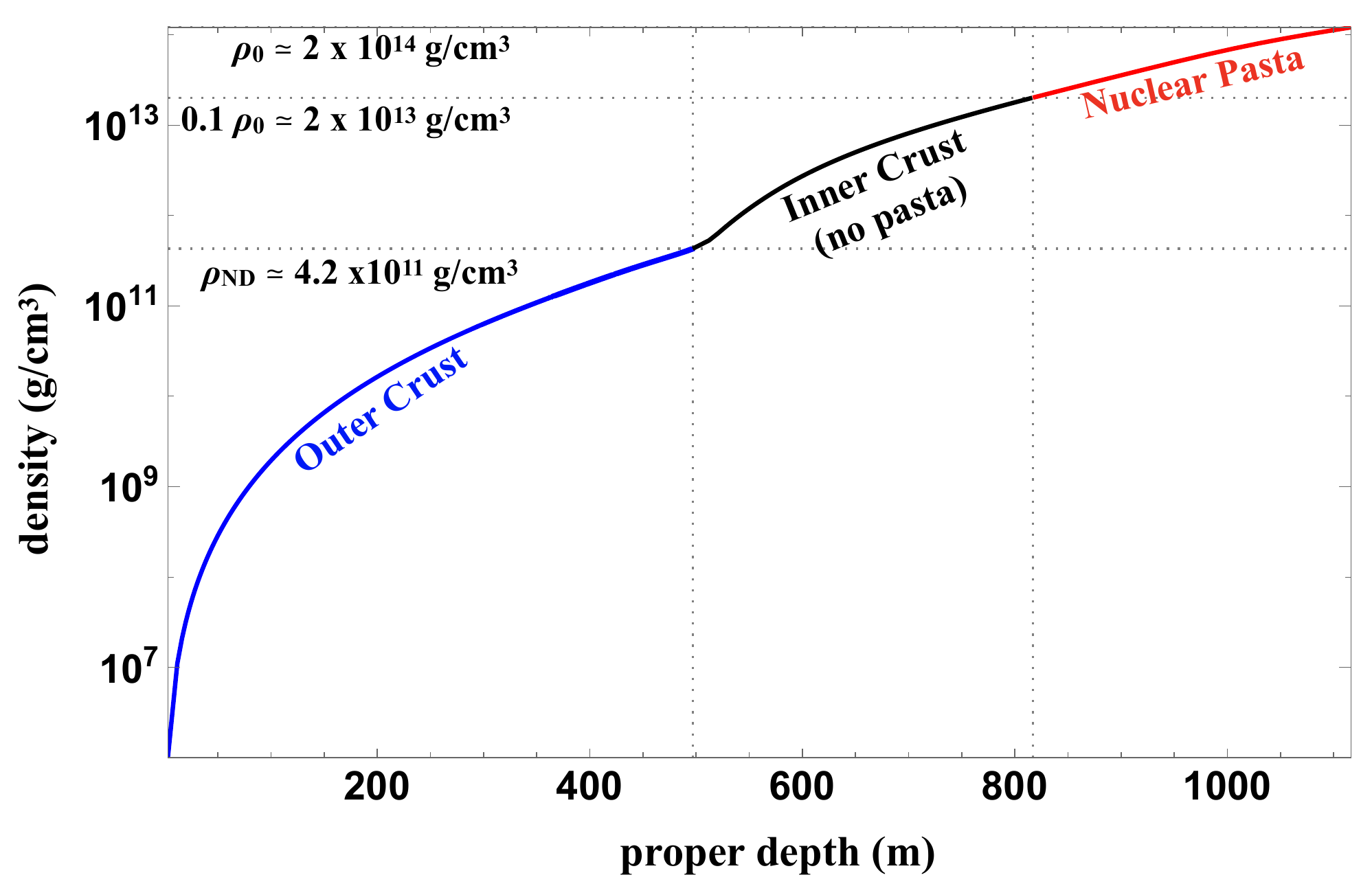} 
\caption{The density of neutron star crust as a function of proper depth, obtained using a benchmark high-density equation of state BsK21 for the benchmark scenario of a 1.8 $M_\odot$ mass, 12.5 km radius neutron star. 
The densities at the transitions between various layers as described in Sec.~\ref{sec:struc} are marked.}
\label{fig:nsinterior2}
\end{figure}

\section{Structure of Neutron Stars}
\label{sec:struc}

The crust of a neutron star, about $1 \ \rm{km}$ thick, spans over 10 orders of magnitude in density and consists of several distinct layers corresponding to different phases of nuclear matter.

The neutron star {\bf atmosphere} and {\bf ocean}, with density $ < 10^4$~g/cm$^3$ and breadth $\mathcal{O}(\mu$m), and made of a thin plasma layer, affect the star's thermal spectrum, and are influenced by the star's magnetic field. 
The {\bf outer crust}  (density $10^4-10^{11}$~g/cm$^3$) is a crystalline lattice of neutron-rich nuclei, along with degenerate free electrons.
The {\bf inner crust} begins below the ``neutron drip line" marked by density $\rho_{\rm ND} \simeq{4.2 \times{10^{11}} \ \rm{g/cm^{3}}}$~\cite{Potekhin:2013qqa}. 
Here it is energetically favourable for neutrons to be liberated from nuclei, and so the inner crust begins as a mixture of a dripped neutron fluid and nuclei in the form of ``clusters" arranged in a lattice.  
Further down, the inner crust density approaches the nuclear saturation point $\rho_0 \simeq {2\times{10^{14}} \ \rm{g/cm^{3}}}$, and intricate competition between nuclear attraction and Coulomb repulsion forms extended non-spherical ``{\bf pasta}" phases of nuclear matter; as the density increases, gnocchi, then spaghetti, and then lasagna pasta phases become more prevalent. 
In the deepest layer of the inner crust there are ``inverted pasta phases'' where nuclear density material predominates over sparser, sub-nuclear density voids. 
This includes bucatini (anti-spaghetti) and swiss cheese (anti-gnocchi) phases.
The crust terminates at densities around $\rho_0$. 
Beyond this lies the {\bf outer core} (densities 0.5$-$2~$\rho_0$) and {\bf inner core} (densities 2$-$10~$\rho_0$). 

Figure~\ref{fig:nsinterior1} shows a schematic of the interior structure of a neutron star, and more closely shows various layers of the inner and outer crust.

Figure~\ref{fig:nsinterior2} shows the density of the crust as a function of the proper depth using a representative  equation of state, BsK21, for representative benchmark neutron star parameters, with mass 1.8 $M_\odot$ and radius 12.5 km.
For further details of the interior of the neutron star, see Appendix~\ref{app:crust}.

We will investigate dark matter scattering on each of these layers. 
We will see in particular that the inner crust and the pasta phases play a key, non-trivial role in dark matter capture, and have important implications for telescope searches for dark matter heating of neutron stars.

\section{Review of Dark Matter Capture and Kinetic Heating}
\label{sec:revcap}

We now provide a brief overview of the dynamics of dark matter capture and kinetic heating in neutron stars.

The steep gravitational potential of a neutron star accelerates non-relativistic dark matter particles falling into it to relativistic speeds. The kinetic energy of the dark matter of mass $\mdm$ at the neutron star surface (in the stellar rest frame) is given by
\begin{equation}
 E_{\rm DM} = \mdm (\gamma-1)~,
 \label{eq:kin-energy1}
\end{equation}
where $\gamma=(1-v^2_{\rm esc})^{-1/2}$, with the escape velocity at the star surface $v_{\rm esc} = \sqrt{2G\Mstar/\Rstar}$, and $\Mstar$ and $\Rstar$ are the mass and radius of the star.
For our study, we adopt the benchmark values
\begin{equation}
\Mstar = 1.8~M_\odot, \ \ \  \Rstar = 12.5~{\rm km},
\label{eq:benchmarkstar}
\end{equation}
so that we have $v_{\rm esc} = 0.65$ and $\gamma$~=~1.32. 
These parameters are chosen to represent realistic and conservative neutron star values; we will see shortly that for neutron star crust capture, {\em less} massive stars tend to capture {\em more} dark matter.
A dark matter particle is captured, {\em i.e.} becomes gravitationally bound to the star, if it loses its halo kinetic energy through one or more scatters with stellar constituents.
The total dark matter mass capture rate is given by~\cite{Goldman:1989nd,Baryakhtar:2017dbj}
\begin{equation}
 \dot{M}_{\rm DM}= \rhoDM v_{\rm halo} \times \pi b^2_{\rm max} \times f,
\label{eq:masscapturerate}
\end{equation}
where $\rhoDM = {0.42 \ \rm{GeV/cm^{3}}}$ and $v_{\rm halo} \simeq 230$~km/s~\cite{Pato:2015dua} are the best-fit halo dark matter density and average speed respectively, and $b_{\rm max}=\gamma\Rstar \times v_{\rm esc}/v_{\rm halo}$ is the maximum impact parameter for dark matter to intersect the neutron star~\cite{Baryakhtar:2017dbj}.
The factor $f$ is the fraction of dark matter particles captured after passing through the neutron star, given by
\begin{equation}
 f={\rm{Min}}\left[1,\frac{\sigma_{\rm T\chi}}{\sigma_{\rm T\chi}^{\rm cap}}\right]~,
\end{equation}
where $\sigma_{\rm T\chi}$ is the cross section for dark matter scattering on a target ($e.g.$, nucleon, lepton) in the star, and 
$\sigma_{\rm T\chi}^{\rm cap}$ is the minimum cross section required for dark matter to capture.
The rate of dark matter kinetic energy transfer to the neutron star is 
\begin{equation}
\dot{\mathcal{E}_k} = \dot{M}_{\rm DM} (\gamma - 1) \times \kappa~,
\label{eq:KEpower}
\end{equation}
where $\kappa$ is the fraction of the dark matter kinetic energy that is deposited into stellar constituents.
This kinetic energy deposited by dark matter could be the predominant heat source for an old neutron star~\cite{Baryakhtar:2017dbj}. 
Over timescales of interest the neutron star will rapidly attain thermal equilibrium, and so $\dot{\mathcal{E}_k}$ is also the rate at which the neutron star radiates heat.
Assuming black body radiation, this luminosity = $4\pi \Rstar^2 \sigma_{\rm SB}T_{\rm eff}^4$ where $\sigma_{\rm SB}$ is the Stefan-Boltzmann constant and $T_{\rm eff}$ is the ``effective" (surface) temperature of the neutron star. 
For a distant observer, $T_\infty = T_{\rm eff}/\gamma$. Thus for the benchmark star parameters given in Eq.~\eqref{eq:benchmarkstar} and $\kappa = 1$ in Eq.~\eqref{eq:KEpower}, we have the dark matter kinetic heating temperature 
\begin{equation}
T_\infty = 1730 \ {\rm K}.
\label{eq:T_signal}
\end{equation}
Without this heating mechanism, an isolated 10$^9$~year-old neutron star would cool down to $\mathcal{O}$(100)~K~\cite{Page:2004fy,Yakovlev:2004iq}. Other than dark matter heating, neutron stars may also be reheated by other astro-nuclear model-dependent internal heating mechanisms~\cite{Gonzalez_2010} and the accretion of interstellar medium, which may be deflected by the star's magnetic fields and preferentially heat its poles~\cite{Baryakhtar:2017dbj}. In the case of extremely fast-rotating (sub-7 ms period) neutron stars and for certain equations of state, rotochemical heating should also be considered~\cite{Hamaguchi:2019oev}. 
However for most neutron stars after 10$^9$ years, all these heating processes (except dark matter kinetic heating) are expected to be negligible.

\section{Dark Matter Capture and Kinetic Heating in Neutron Star Crusts}
\label{sec:capture}

This section implements a neutron star crust framework, then describes in detail how dark matter may be captured by each crustal layer.

\subsection{Framework for Neutron Star Crusts}

First, we will take the trajectory of dark matter through the crust to be a straight line that passes through the center of the star. 
This is a conservative choice, since any other trajectory implies a longer path for dark matter through the crust and hence more scattering targets along its path. 
The optical depth of a dark matter particle in the crust is
\begin{equation}
 \tau_{\rm DM} = \int_{\rm crust}{n_{\rm T} \sigma_{\rm T\chi}}  dz~,
 \label{eq:scatt-nmbr}
\end{equation}
where 
$n_{\rm T}$ is the number density of scattering targets, 
$\sigma_{\rm T\chi}$ is the cross section for dark matter scattering on a target, 
and the integration variable $z=(\Rstar-r)\gamma$ is the proper distance traveled by dark matter in the crust, where $r$ is the Schwarzschild radial co-ordinate.
Note that $z$ integrates over every crustal layer twice, to account for dark matter both entering and exiting the star.
The integral in Eq.~\eqref{eq:scatt-nmbr} can be conveniently recast in terms of the crust density by using the light-and-thin crust approximation~~\cite{Chamel2008,Haensel:2007yy}, valid for neutron stars of typical mass and radius such as the one we have considered. 
Since the mass and thickness of the crust is $\lesssim$ 10\% of the star's total mass and radius, the equation for hydrostatic equilibrium in Eq.~\eqref{eq:tov} can be simplified to yield~\cite{Chamel2008,Haensel:2007yy}
\begin{equation}
 \frac{dP}{dz}=g_{s}\rho,
 \label{eq:depth-thincrust}
\end{equation}
where $P$ is the pressure and $g_{s}=G\Mstar\gamma/\Rstar^2$ is the acceleration due to gravity at the star surface.
If the equation of state, {\em i.e.} $P$ as a function of the density $\rho$, is known, it is helpful to use Eq.~\eqref{eq:depth-thincrust} to rewrite Eq.~\eqref{eq:scatt-nmbr} as
\begin{equation}
 \tau_{\rm DM} = \frac{1}{g_{s}} \int_{\rm crust}{n_{\rm T} \sigma_{\rm T\chi}}  \frac{1}{\rho} \frac{dP}{d\rho} d\rho~,
 \label{eq:scatt-nmbr2}
\end{equation}
where $dP/d\rho$ may in practice be computed from analytical representations of the equation of state; see Appendix~\ref{app:crust}.

We will assume that scattering is isotropic in the dark matter-target center-of-momentum frame, which results in a flat spectrum of energy transfers:
\begin{equation}
\frac{d\sigma}{d\Delta E} = \frac{\sigma_{\rm T\chi}}{\Delta E_{\rm max}}~,
\label{eq:recoilEspectrum}
\end{equation}
where $\Delta E_{\rm max}$ is the maximum energy transferred.
For this spectrum the average energy transferred = $\Delta E_{\rm max}/2$; for  dark matter recoiling against a single nucleon this is~\cite{Baryakhtar:2017dbj}
\begin{equation}
 \Delta{E}_{\rm ave}=\frac{m_{\rm n}\mdm^{2}\gamma^{2}v_{\rm esc}^{2}}{(m_{\rm n}^{2}+\mdm^{2}+2\gamma{m_{\rm n}\mdm)}}.
 \label{eq:Eloss}
\end{equation}
This transferred energy exceeds the dark matter halo kinetic energy $ \mdm v^2_{\rm halo}/2$ for $\mdm \lesssim 10^6~\gev$.
Consequently, a single scatter in the crust suffices to capture dark matter in this mass range, which will be the focus of this work.
For higher dark matter masses multiple scatters will be required for capture; a careful treatment of dark matter capture via multiple scattering in the crust is left for future study.

Once a dark matter particle is gravitationally captured, it oscillates around the star in an orbit that rapidly shrinks as it transits the crust multiple times, until it is completely contained within the star.
This stage of thermalization occurs on a short timescale; see Appendix~\ref{app:thermstage1} for a full general-relativistic derivation of this.
The net energy that dark matter deposits in the crust during this process is 
\begin{equation}  
 \Delta{E}_{\rm orbit} = (\mdm + \mdm v^2_{\rm halo}/2) - (\mdm / \gamma)~,
 \label{eq:DeltaEcrust}
\end{equation}
where the term in the first parentheses is the energy of the particle far away from the star, and the term in the second parentheses is its gravitational binding energy at the star surface.
Eqs.~\eqref{eq:kin-energy1} and \eqref{eq:DeltaEcrust} imply that the fraction of dark matter kinetic energy deposited in the crust $\Delta{E}_{\rm orbit} / E_{\rm DM}$ (with $v^2_{\rm halo} \ll 1$ neglected) is given by
\begin{equation}
\kappa_{\rm crust} = \frac{1}{\gamma}~.
\end{equation}
The neutron star crust then radiates heat from the surface at a rate equal to the kinetic heating rate, with $\kappa = \kappa_{\rm crust}$ in Eq.~\eqref{eq:KEpower}.
The dark matter heating temperature obtained from this luminosity is
\begin{equation}
T_\infty^{\rm crust} = 1620~{\rm K}.
\label{eq:Tsignalcrust}
\end{equation}
This is again much greater than the $\mathcal{O}$(100)~K temperatures expected without dark matter heating. In an optimal scenario, it may be possible to discern between crust vs. core heating using a population of old neutron stars; see Appendix~\ref{app:discern} for details.

In the next few sub-sections we investigate in detail dark matter scattering on the components of the various layers of the neutron star crust mentioned in Sec.~\ref{sec:struc}.
We then estimate the sensitivity to dark matter capture of each of these layers.

\subsection{Scattering with the Outer Crust}

In the outer crust, where nuclei form an ordered lattice, dark matter scattering on individual nucleons in the nuclei can result in dark matter capture. 
The nuclear binding energy per nucleon is roughly the same as that measured at laboratory densities: $E_{\rm bind} \lesssim 10 \ \mev$~\cite{RocaMaza:2008ja}, which decreases deeper into the crust, where nuclei become heavier and richer in neutrons. 
For $\mdm \gtrsim 10 \ \mev$, dark matter would scatter on these weakly bound nucleons by transferring energies greater than $ E_{\rm bind}$. 
To calculate the \textit{per-nucleon} capture cross section in the outer crust, we integrate Eq.~\eqref{eq:scatt-nmbr2} over the outer crust densities, which are from 10$^6$~g/cm$^3$ to $\rho_{\rm ND} \simeq 4.2 \times 10^{11}$~g/cm$^3$.

The \textit{per-nuclear} cross section for dark matter scattering on a nuclear species T is given by
\begin{equation}
\sigma_{\rm T\chi}(q) = \left(\frac{\mu_{\rm T\chi}}{\mu_{\rm n\chi}}\right)^{2} A^2 F^2(q)S_{\rm T}(q)\sigma_{\rm n\chi},
 \label{eq:nuclearcs}
\end{equation}

where $\mu_{\rm T\chi}$ ($\mu_{\rm n\chi}$) is the nucleus-dark matter (nucleon-dark matter) reduced mass, 
$q$ is the momentum transferred to the nucleus,
$F(q)$ is the Helm form factor~\cite{Helm:1956zz,LEWIN199687}, and 
$S_{\rm T}(q)$ is the static structure function per nucleus~\cite{kittel1987quantum}, which has been computed for dense astrophysical plasmas~\cite{Baiko:1998xk,2000PhRvE..61.1912B}. 
To calculate the per-nuclear capture cross section in the outer crust, we substitute Eq.~(\ref{eq:nuclearcs}) into Eq.~\eqref{eq:scatt-nmbr2} and integrate over the outer crust densities, which are from 10$^6$~g/cm$^3$ to $\rho_{\rm ND} \simeq 4.2 \times 10^{11}$~g/cm$^3$.
 Here the number density in Eq.~\eqref{eq:scatt-nmbr2} is simply $n(\rho)=\rho/m_{n}$. 

Scattering on the \textit{nuclei} in the outer crust is unlikely to result in dark matter capture, due to the interplay of two factors:

\begin{enumerate}

 \item When the de Broglie wavelength $q^{-1}$ is \textit{less than} the nuclear radius $r_a$, there is loss of coherence over a nucleus, captured by the form factor $F(q)$ which suppresses $\sigma_{\rm T\chi}$.
 
 \item When the de Broglie wavelength $q^{-1}$ is \textit{greater than} the inter-nuclear separation $a$, there is loss of coherence among the relative amplitudes of dark matter scattering on {\em multiple} nuclei, captured by   $S_{\rm T}(q)$, which suppresses $\sigma_{\rm T\chi}$ \cite{Haensel:2007yy}.
  
\end{enumerate}

This can be seen explicitly by considering the relevant parameters for neutron stars:
Nuclear radii in the outer crust are approximated by the standard formula $r_a\simeq{(1.25 \ {\rm{fm}} \ A^{1/3})}\gtrsim 0.04 \ \gev^{-1}$ \cite{Pethick:1995di}. 
The inter-nuclear separation for nuclear species with mass $m_{\rm T}$ in the outer crust is $a \sim (\rho/ m_{\rm T})^{1/3}\gtrsim 0.02 \ \gev^{-1}$. 
The average momentum transfer $\langle q \rangle\sim \mdm \gamma v_{\rm esc}$ for $\mdm \ll m_{\rm T}$, where $m_{\rm T} \sim 50-200 \ \gev$; see Fig.~\ref{fig:za}. 
Comparing the above distance scales to $\langle q \rangle^{-1}$, we find that for $\mdm \gtrsim 30 \ \mev$, the scattering cross section is suppressed by $F(q)$, whereas for lighter dark matter, it is suppressed by $S_{\rm T}(q)$. 
For $\mdm \gtrsim  m_{\rm T}$, $\langle q \rangle\sim  m_{\rm T}\gamma v_{\rm esc} \gtrsim r_{a}^{-1}$, so that the cross section is suppressed again by $F(q)$. 
Thus, the product $F^{2}(q)S_{a}(q)$ in Eq.~\eqref{eq:nuclearcs} suppresses $\sigma_{\rm T\chi}$ for practically all dark matter masses. 
Therefore, in the outer crust, dark matter capture occurs mostly via scattering on individual nucleons, and not via scattering on nuclei.

\subsection{Scattering with the Inner Crust}

In the inner crust, dark matter capture can occur via scattering in two different ways: with nucleons weakly bound in nuclei, and with superfluid neutrons.

For interactions with nucleons, the greater densities in the inner crust imply that the binding energy per nucleon is even smaller than in the outer crust.  Once again, therefore, dark matter scattering on weakly bound nucleons can result in capture for $\mdm \gtrsim 10 \ \mev$.
To calculate the capture cross section via scattering on nucleons in the inner crust, we again integrate Eq.~\eqref{eq:scatt-nmbr2}, but now integrate over inner crust densities, which are from $\rho_{\rm ND}$ to 0.1~$\rho_0$.
We will expect the capture cross section here to be $1-2$ orders of magnitude smaller than the outer crust, due to the greater densities of the inner crust, which make it optically thicker. We note that like the outer crust, in the inner crust dark matter scattering with whole nuclei is suppressed\footnote{This can be seen by applying the outer crust arguments in the previous section with parameters for nuclei in the inner crust, where nuclear radii are instead well-approximated by $r\propto Z^{1/3}$.}.

In the inner crust, dark matter could also be captured by depositing energy on the free/dripped neutrons, which form a superfluid phase at the low stellar temperatures ($\mathcal{O}(10^3$~K)) that we consider.
This phase has an energy gap of $\Delta_{F} \simeq 1 \ \mev$~\cite{Cao:2006gq}; therefore for $\mdm \gtrsim$ 10 MeV, scattering  on individual nucleons can unquestionably proceed. One might also consider dark matter capture through the breaking of neutron Cooper pairs~\cite{Bardeen:1957mv}, formed by spin-paired neutrons in superfluid neutron matter in the inner crust~\cite{Kundu:2004mz,Bertoni:2013bsa,Bertoni:2014soa}. 
The pairing energies are expected to be $\sim 0.01-0.1~{\rm MeV}$, although the dynamics and energy gaps for Cooper pairs in superfluid neutron matter are uncertain as they depend on many factors such as polarization contributions to the spin interactions of neutrons~\cite{Schwenk:2003bc}. 
This is a topic of active research~\cite{Gezerlis:2014efa}, and so to be conservative we leave a detailed treatment of dark matter capture via breaking inner crust neutron Cooper pairs to future work.

For dark matter masses much less than the energy or pairing gaps, dark matter can still deposit energy by exciting neutron superfluid quasi-particles such as phonons.
Prior work specifically considered light dark matter's effect on the thermal conductivity of young, hot ($T \gtrsim 10^7$ K) neutron stars \cite{Cermeno:2016olb}. 
Here we will only provide a rough estimate of dark matter capture by phonon modes.
Single-phonon emission in the low momentum regime with a linear dispersion relation is described by a static structure function, which relates the per-nucleon cross section to the phonon excitation cross-section~\cite{Bertoni:2013bsa,Cermeno:2016olb}:
\begin{equation}
 S_{\rm phonon}(q) = \frac{q}{2m_{\rm n}c_{\rm s}},
\label{eq:Sqphonon}
\end{equation}
where $c_{\rm s}$ is the speed of the superfluid phonon which at leading order is the neutron Fermi speed equal to $(3 \pi^{2}n_{\rm n})^{1/3}/m_{\rm n}$, where $\simeq~$0.04~$c$ at the neutron drip density. 
We note that this linear regime, valid for sufficiently low momentum dark matter, does not account for possible contributions from higher energy excitations ($e.g.$ maxons, rotons). 
For this reason, we will only consider low mass (and correspondingly low momentum) dark matter parameter space.
Put another way, we require that $S_{\rm phonon}(q)$ in Eq.~\eqref{eq:Sqphonon} is much less than unity, which for $q \simeq \gamma v_{\rm esc} \mdm \simeq \mdm$ is valid for $\mdm \lesssim 10^{-3} ~{\rm GeV}$. 
Future work could be undertaken to compute higher mass dark matter capture via phonon excitations, although this will depend on the specifics of the crust model. 
Finally, we note that we have been conservative in our choice of sound speed $c_{\rm s} \simeq 0.04$~$c$.
While we have neglected possible mixing between the superfluid phonons and nuclear lattice phonons, including this effect would tend to decrease the sound speed of the lowest lying modes by up to a factor of 4~\cite{Chamel:2012ix}, and result in capture cross sections proportionally smaller. We leave the full treatment of these effects to a future study.

Where Eq.~\eqref{eq:Sqphonon} is valid, the energy deposited by dark matter via exciting phonons is $\sim q c_{\rm s}$.
For dark matter masses much less than the neutron mass, the momentum transfer $q$ is about $\mdm \gamma v_{\rm esc}$. 
This means that this deposited energy is much greater than the dark matter kinetic energy $\gamma \mdm v^2_{\rm halo}/2$, and so dark matter will  be captured via excitation of a single phonon.
Therefore, 
accounting for the factor $S_{\rm phonon}(q)$, 
the capture cross section for scattering via phonon emission is equal to $10^{-39} \ {\rm{cm^{2}}}~(1~\mev/ \mdm)$ for dark matter masses less than the energy gap $\Delta_F = 1 \ \mev$.

\subsection{Scattering with Nuclear Pasta}

We now discuss our ``\textit{penne trap}''~\cite{Brown:1985rh} for dark matter. 
The cross section for dark matter scattering on nuclear pasta is given by
\begin{equation}
 \sigma_{\rm pasta}(q)=S_{\rm pasta}(q) \ \sigma_{\rm n \chi}~,
 \label{eq:pasta1}
\end{equation}
where $S_{\rm pasta}(q)$ is a structure or response function accounting for correlations between {\em nucleons} in the pasta. 
Here we have neglected the per-proton cross section on the right hand side of the equation since the contribution to $S_{\rm pasta}(q)$ from neutron-neutron correlations will dominate over neutron-proton and proton-proton correlations as neutrons outnumber protons in nuclear pasta by $\mathcal{O}$(100); see for instance Fig.~\ref{fig:za}.

Nuclear pasta has important implications for the properties of neutron stars and core collapse supernovae. For example, neutrino-pasta interactions impact neutrino opacity in supernovae, and electron-pasta interactions impact properties such as the neutron star shear viscosity, thermal and electrical conductivity. 
As such, the static response functions of neutrino and electron scattering with pasta have been previously calculated in several simulations~\cite{Magierski:2001ud,Gogelein:2007pb,Sonoda:2007sx,Iida:2001xy,Watanabe:2006qf,Schneider:2013dwa,daSilvaSchneider:2018yby,Caplan:2018gkr}. 
We apply the Quantum Molecular Dynamics (QMD) simulation results~\cite{Nandi:2017aqq}, as these are relevant for our stellar temperatures and dark matter momentum transfers. 

The QMD results of Ref.~\cite{Nandi:2017aqq} provide the low-temperature structure factor for several densities (and so several pasta phases) with a proton fraction of $Y_p= 0.3$, and only the gnocchi phase when the proton fraction is $Y_p = 0.1$. We note that the proton fraction of beta-equilibrated matter in a cold inner crust is about 0.1, which is the type of crust relevant for our calculation. However, as we consider only the neutron-neutron correlations, a $0.1-0.3$ variation in proton fraction is expected to yield comparable results. As the gnocchi scattering structure factor is provided at both densities, we have checked that dark matter-gnocchi scattering cross sections with 0.1 and 0.3 proton fractions produce the same results. Given this, we expect that lasagna and spaghetti scattering will also be similar for 0.1 and 0.3 proton fractions, and so use the available $Y_p=0.3$ proton fraction results for those pasta phases.

In any case, it is important to note that the treatment of pasta structure factors only significantly impacts the nucleon scattering enhancement visible in Fig.~\ref{fig:bounds1} for dark matter masses $0.1-1$ GeV. This is because for most of the parameter space, capture occurs through inelastic ejection of nucleons out of nuclear pasta; this process is insensitive to the coherent pasta scattering enhancement. Ideally, we would sample several datasets to examine the variations in this coherent pasta-scattering enhancement, but as these datasets are not available, we provide estimates here. Neglecting this pasta coherent enhancement altogether would simply result in no dip in the nucleon pasta sensitivity curve around DM masses $0.1-1$ GeV.

The QMD simulations capture the coherence effects of interacting with the whole pasta structure. 
Thus, $\sigma_{\rm pasta}$ must be interpreted as a {\em per-nucleon} cross section.

For sub-nuclear densities, the response $S_{\rm pasta}(q)$ can be roughly divided into three regimes in $q$~\cite{Nandi:2017aqq}: 

\begin{enumerate}

 \item  $q\lesssim$~40~MeV: The response function $S_{\rm pasta}(q)$ $\ll 1$, as the scattering is incoherent due to density fluctuations, which results in destructive interference of scattering amplitudes.
In this region it is given by $S_{\rm pasta} \rightarrow T(\partial\rho/\partial P)_{T}$, where $T$ is the pasta temperature.
For example, at $T = 1000$~K, $S_{\rm pasta}(q)$ is $\mathcal{O}(10^{-6})$. 

 \item  $q$~$\simeq$~60~MeV: The response function $S_{\rm pasta}(q)$ peaks, due to coherent enhancement from scattering on multiple nucleons. 

 \item  $q\gtrsim$~200~MeV: The response function $S_{\rm pasta}(q) \rightarrow 1$, corresponding to quasi-elastic scattering on weakly bound individual neutrons. 
\end{enumerate}

To calculate the dark matter-pasta scattering sensitivities, we modify Eq.~\eqref{eq:scatt-nmbr2} for the pasta layers as
\begin{equation}
  \tau_{\rm DM}  =\frac{\sigma_{\rm n\chi}}{g_{s}}\int_{\rm pasta} \langle S_{\rm pasta}(q) \rangle_{q} \ \frac{n_{\rm n}(\rho)}{\rho}\frac{dP}{d\rho} d\rho~,
 \label{eq:Nscatterspasta}
\end{equation}
where $n_{\rm n}$ is the nucleon number density. 
The $q$-averaged structure function is given by 
\begin{equation}
\langle S_{\rm pasta}(q) \rangle_{q} = \frac{1}{q_{\rm max}}\int_{0}^{q_{\rm max}} S_{\rm pasta}(q) \ dq~,
\end{equation}
with the maximum momentum transfer
\begin{equation}
 q_{\rm max}=\sqrt{\frac{4\mu_{\rm n\chi}\mdm^{2}m_{\rm n}\gamma^{2} v_{\rm esc}^{2}}{\mdm^{2}+m_{\rm n}^{2}+2\mdm m_{\rm n}\gamma}}~.
\end{equation}

We then integrate Eq.~\eqref{eq:Nscatterspasta} over the pasta layer densities, which extend from 0.1~$\rho_0$ to 0.3~$\rho_0$.
This range is a conservative choice, as various pasta structures are expected to appear near the saturation density $\rho_0$, which may contribute to dark matter capture via quasi-elastic scattering on individual nucleons.
Thus our range of integration includes nuclear spheres (gnocchi), cylinders (spaghetti) and slabs (lasagna), but not cylindrical (bucatini) and spherical (Swiss cheese) voids~\cite{Sonoda:2007sx,Watanabe:2006qf}.

\subsection{Results and Discussion}
\label{subsec:captureremarks}

\subsubsection{Optical depths}

We first present the optical depth of each layer by integrating over the appropriate density ranges as laid out in Eq.~\eqref{eq:scatt-nmbr2}:

\begin{eqnarray}
\nonumber \tau_{\rm DM}^{\rm outer}&\simeq& \left(\frac{\sigma_{\rm n\chi}}{2\times 10^{-40} \ \rm{cm^{2}}}\right)~, \\
\nonumber \tau_{\rm DM}^{\rm inner} &\simeq& \left(\frac{\sigma_{\rm n\chi}}{7\times 10^{-42} \ \rm{cm^{2}}}\right)~, \\
\tau_{\rm DM}^{\rm pasta} &\simeq& \left(\frac{\sigma_{\rm n\chi}}{10^{-42} \ \rm{cm^{2}}}\right)~.
\end{eqnarray}

We have normalized the per-nucleon cross sections so that the optical depth is unity, which corresponds to the spin-independent cross section required for dark matter capture in each layer, for dark matter masses $10-10^6$ GeV, as seen in Fig.~\ref{fig:bounds1}.

\subsubsection{Layer by Layer Results}

\begin{figure*}[t] 
\includegraphics[width=.77\textwidth]{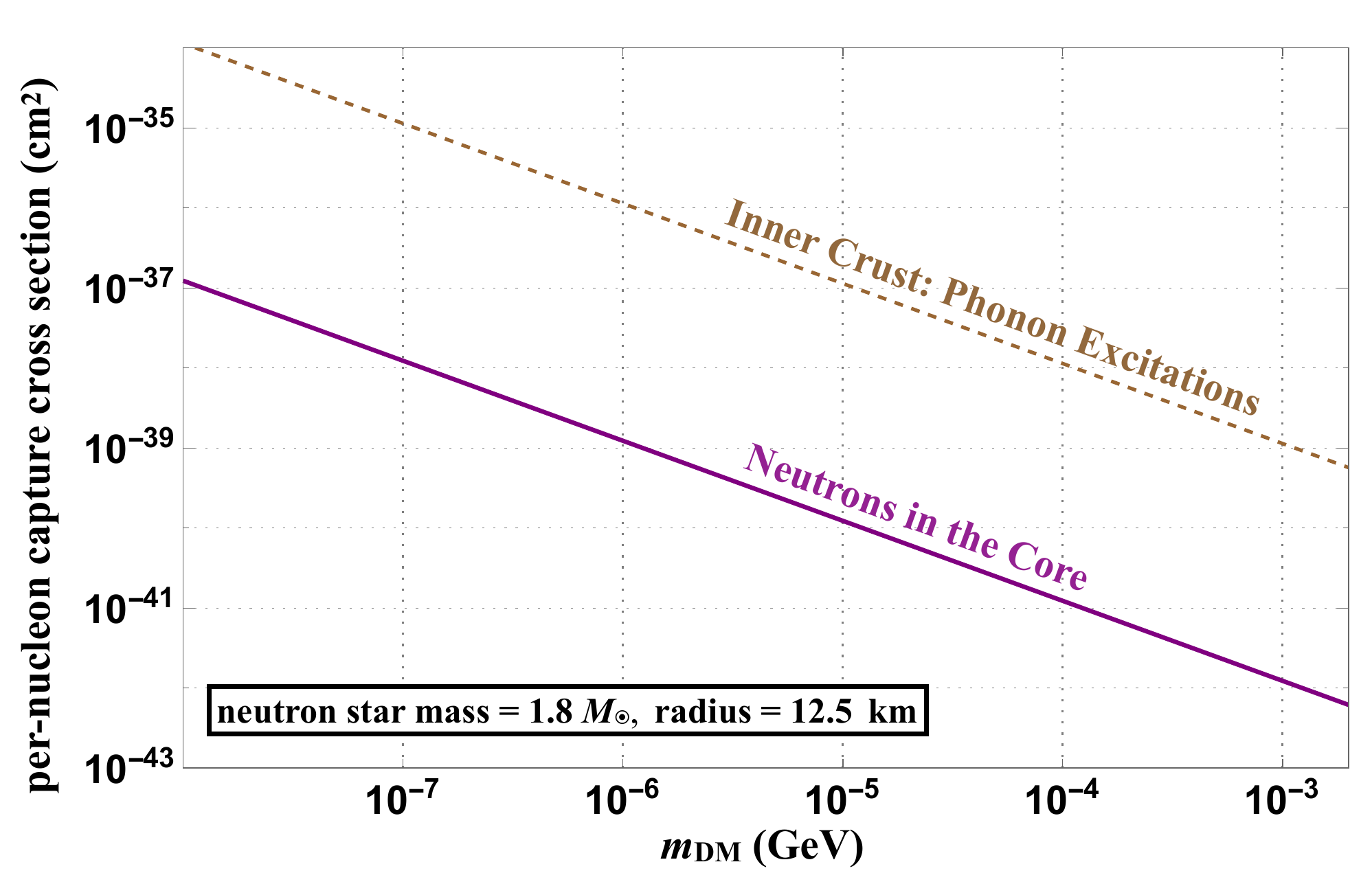}\\
\includegraphics[width=.77\textwidth]{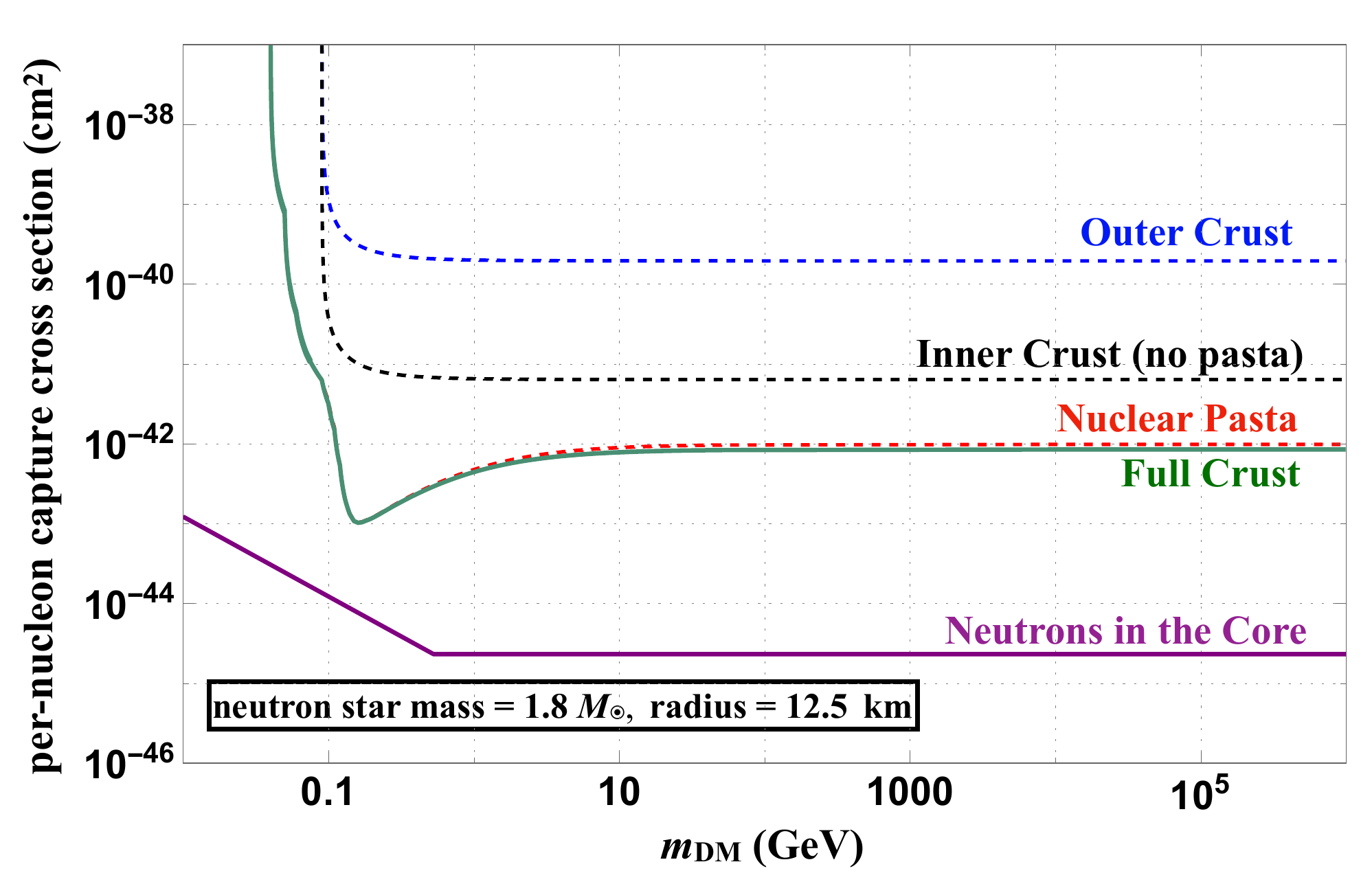} \\
\caption{Dark matter-nucleon scattering cross section for capture by neutron star layers as a function of dark matter mass $\mdm$, for a neutron star of mass 1.8~$\Mstar$ and radius 12.5~km. \textbf{Top:}
For $\mdm \lesssim$ MeV, dark matter captures by scattering on superfluid neutrons in the inner crust and exciting collective modes (phonons) via low momentum transfers.
\textbf{Bottom:} For $\mdm \gtrsim$ 100 MeV, dark matter captures by exciting weakly bound nucleons in the nuclei of the outer and inner crust, as well as in nuclear pasta at the bottom of the inner crust. 
In the inner crust, scattering on free neutrons also contributes to capture.
Also shown is an estimate of the capture cross sections for dark matter scattering with the neutron star core assuming degenerate neutron matter; however the dynamics of the core are not reliably known.}
\label{fig:bounds1}
\end{figure*}

Figure~\ref{fig:bounds1} summarizes our results for dark matter capture in neutron star crusts. 
In the top panel, we have plotted the per-nucleon capture cross section for $\mdm \lesssim$~1~MeV, and in the bottom panel where $\mdm \gtrsim$~10~MeV, we have plotted the same accounting for scattering in all the layers of the crust. Our results can be summarized as follows:

\begin{itemize}
 \item \underline{$\mdm \gtrsim$~90~MeV:}
 
This is the mass range in which capture could occur in all crustal layers.
In general, the sensitivity to dark matter capture is driven by the pasta layers. 
This is mainly because, even though the pasta layers are very thin, their enormous densities give them enormous optical thickness. 

For 100~MeV~$\lesssim \mdm \lesssim$~1~GeV, dark matter-pasta scattering is coherently enhanced, which improves capture sensitivity by as much as an order of magnitude.

 For 90~MeV~$\lesssim \mdm \lesssim$~400~MeV, a fraction of scatters can occur with energy transfer greater than 10 MeV (the nuclear binding energy of nucleons), as determined from the spectrum in Eq.~\eqref{eq:recoilEspectrum}.
 Thus the capture cross section decreases with $\mdm$ in this mass range.
 Finally, for $\mdm \gtrsim$~400~MeV, almost all scatters occur with energy transfer greater than the nuclear binding energy 10 MeV, so that the capture cross sections of the outer crust and pasta-free inner crust are $\mdm$-independent in this mass range.

\item \underline{10~MeV~$\lesssim \mdm \lesssim$~90~MeV:}

Dark matter capture by pasta is suppressed by incoherent scattering due to density fluctuations, so that the net capture sensitivity is driven by dark matter scattering on nucleons in the inner crust.
The outer crust sensitivity is weaker than these layers' by $1-2$ orders of magnitude simply because it is much less dense.

In this mass range, there are no scatters with energy transfer greater than the nuclear binding energy 10 MeV, as seen from Eqs.~\eqref{eq:recoilEspectrum} and \eqref{eq:Eloss}.
 Therefore, capture via scattering on weakly bound nucleons shuts off, seen as vertical lines in the bottom panel of Fig.~\ref{fig:bounds1}.

\item  \underline{10 eV~$\lesssim \mdm \lesssim$~1~MeV:} 

Sensitivity to dark matter capture in the crust is driven solely by the excitation of collective modes (phonons) in the neutron superfluid phase of the inner crust.
In this region, both the crust and core sensitivities scale as $\mdm^{-1}$ since the cross sections for both phonon excitation in the inner crust (Eq.~\eqref{eq:Sqphonon}) and neutron excitation in the degenerate core contain a factor of the momentum transfer proportional to the dark matter mass. Our estimates of capture in this region are valid down to $\mdm \simeq 10$~eV, for which dark matter evaporation may become important for a 1000 K-hot neutron star~\cite{Garani:2018kkd}.

\end{itemize}

Note that we do not show sensitivities in the mass range 1~MeV~$\lesssim \mdm \lesssim$~10~MeV. The capture calculations in this range are much more complicated than our current, conservative treatment; this is a region where the dark matter masses are greater than the superfluid energy gap $\simeq$~1~MeV, and less than the per-nucleon binding energy $\simeq$~10~MeV. We leave a detailed treatment of this mass range for future investigation.

Remarkably, we see from the dark matter-core scattering bounds in Fig.~\ref{fig:bounds1}, that the cross section sensitivity of neutron star crusts to dark matter kinetic heating is at most only 3 orders of magnitude weaker than that of the neutron star core.
The latter scales as $\mdm^{-1}$ for $\mdm \lesssim$~1~GeV due to Pauli-blocking of degenerate neutrons~\cite{Baryakhtar:2017dbj}. Indeed, because of a structure factor enhancement in the pasta phase, the crust sensitivity is only a factor of 10 below the core sensitivity for $\mdm \simeq$~100~MeV.

The separation of 3 orders of magnitude between the sensitivities of the pasta (without coherent enhancement) and core layers can be understood simply from their respective optical depths. 
From Fig.~\ref{fig:nsinterior2} and the discussion in Sec.~\ref{sec:struc}, we see that the inner crust is $\mathcal{O}$(10) less dense and $\mathcal{O}$(10$^2$) thinner than the core, so that its optical depth is $\mathcal{O}$(10$^3$) smaller.
Similarly we see from Fig.~\ref{fig:nsinterior2} that the inner crust excluding pasta is $\mathcal{O}$(10) less dense than, but about as thick as the pasta layer; therefore the inner crust optical depth is $\mathcal{O}$(10) smaller, which is reflected in Fig.~\ref{fig:bounds1}. 
Further, the densest part of the outer crust is $1-2$ orders of magnitude less dense than the inner crust on average, while of comparable thickness; therefore its optical depth (and hence cross section sensitivity in Fig.~\ref{fig:bounds1} is $1-2$ orders of magnitude smaller.

It is important to keep in mind that any of the crust layers could be responsible for the capture of dark matter, due to the order in which it would encounter the layers in its first transit through the crust. 
Thus if the dark matter-nucleon cross section $\sigma_{\rm n \chi}$ happens to exceed twice the outer crust capture cross section, then dark matter capture is effected by the outer crust;
if  $\sigma_{\rm n \chi}$ is between twice the capture cross sections of the outer crust and a deeper layer, capture is effected by that deeper layer. 

It is also important to note that our knowledge of neutron star structure becomes more theoretical and less empirical deeper into the star (see Appendix~\ref{app:crust}). Consequently, our cross section sensitivities are more robust for shallower layers, so that the outer crust sensitivity at about $10^{-40}$~cm$^2$ is the most robust.
This is simply because we understand less well how matter should behave at higher densities and pressures. For example, while almost all EoS predict a pasta phase, there are uncertainties in the types of pasta present, and the density at which they appear~\cite{Pearson:2018tkr}. 

Here we have used the pasta structure function as given in Ref.~\cite{Nandi:2017aqq}, which computed it for neutron stars in the low temperature regime relevant in this study.  
Importantly, even in a scenario with no pasta structure function enhancement, there would still be densely packed nuclear material, whose nucleons dark matter would still scatter on quasi-elastically.
 We emphasize that the only region in which the pasta uncertainties are relevant is for dark matter masses of about $0.1-1$ GeV, where scattering is enhanced by the structure function. 
Neglecting pasta altogether would simply remove the dip in sensitivity around these masses,  evident in Fig.~\ref{fig:bounds1}. 
This would weaken the predicted dark matter-nucleon scattering sensitivity in the $0.1-1$ GeV mass range by less than an order of magnitude. 
For dark matter masses greater than a GeV, scattering on the nuclear pasta layers occurs predominantly through inelastic scattering with nucleons, as given by Eq.~\eqref{eq:inelasticxs}, which does not depend on the nuclear pasta structure function.

The results we have presented have an important consequence for weakly interacting (WIMP) dark matter candidates: as they have electroweak-sized scattering cross sections of about $10^{-39}$~cm$^2$ and GeV$-$TeV masses, they are \textit{guaranteed to capture} in neutron stars irrespective of the composition of the interior below the outer crust.
We will revisit this aspect of capture in the context of inelastic dark matter, with a case study of supersymmetric Higgsino dark matter in Sec.~\ref{subsec:Higgsino}.

\begin{figure*}[t] 
\includegraphics[width=.75\textwidth]{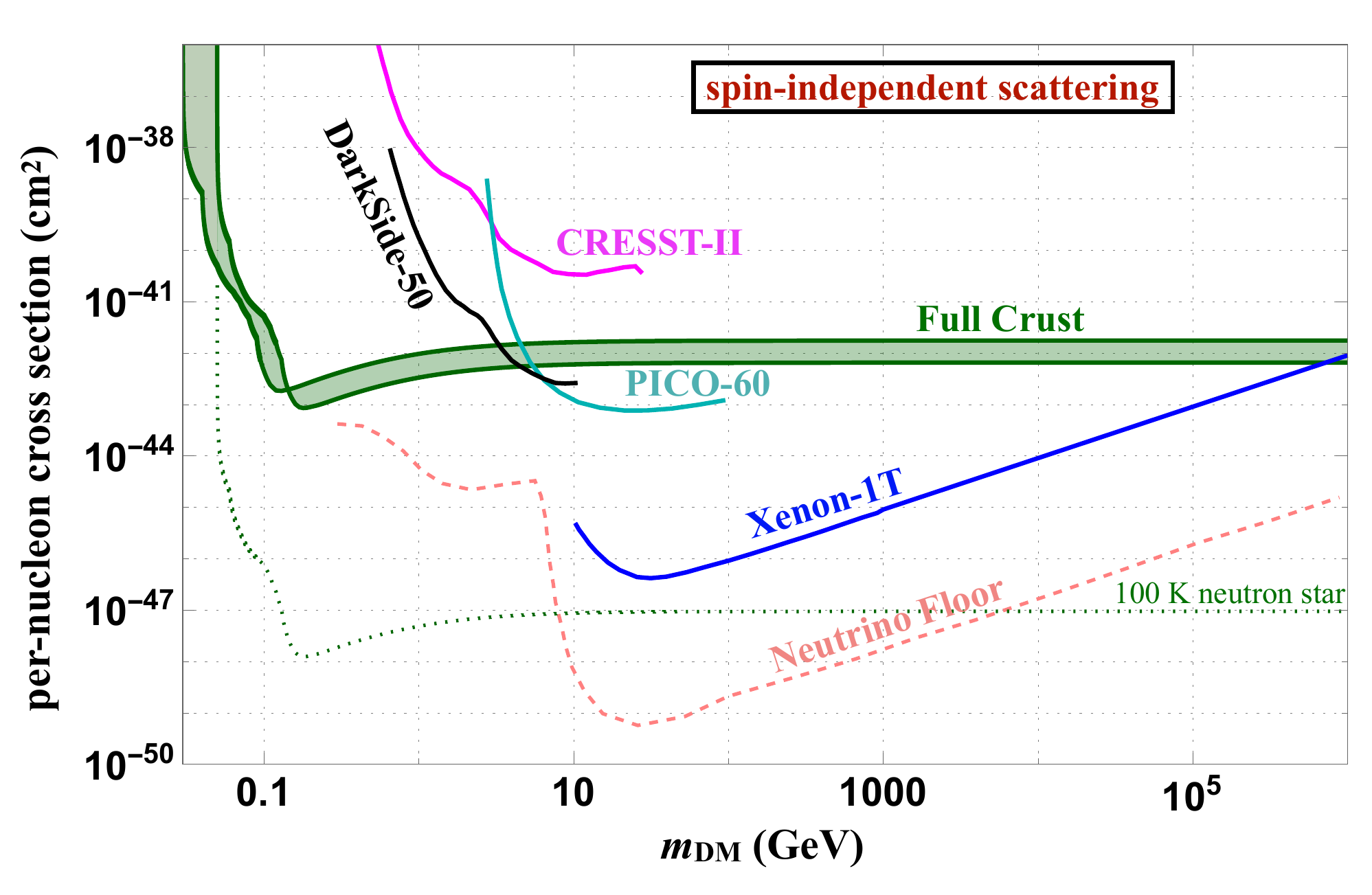}\\
\includegraphics[width=.75\textwidth]{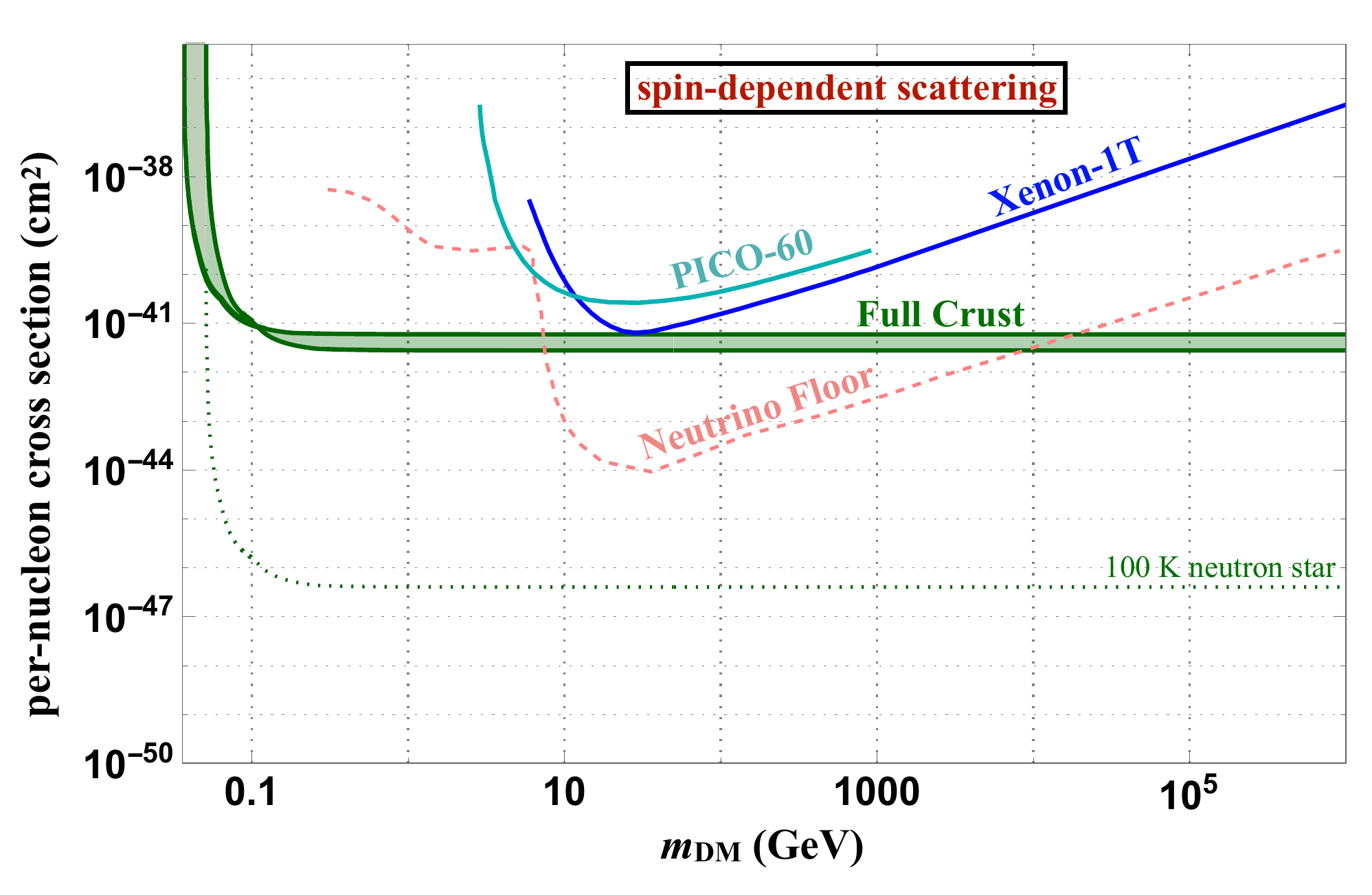} \\
\caption{Comparison of net per-nucleon capture cross sections of neutron star crusts with constraints from underground dark matter direct detection searches, for spin-independent ({\bf \em top}) and spin-dependent ({\bf \em bottom}) scattering.
Also shown are the xenon neutrino floors as a rough estimate of the ultimate sensitivity of direct detection experiments \cite{Billard:2013qya,Ruppin:2014bra}.
The ``Full Crust'' shaded band spans the range of capture cross sections obtained by varying key properties of neutron stars, with the weakest sensitivity provided by \{mass, radius, rotational frequency, high-density equation of state\} =
\{2.3~$M_\odot$, 11.5 km, 716 Hz, BsK21\}, and the strongest sensitivity by 
\{1.4~$M_\odot$, 11.75 km, 0 Hz, BsK20\}.``Full Crust'' shows the cross-section for which all dark matter passing through the neutron star crust is captured, resulting in kinetic heating to $\sim 1620$ K; the cross-section for which DM kinetic heating leads to a ``100 K neutron star'' is also indicated. This is roughly the full optimistic reach.
}
\label{fig:DD}
\end{figure*}

\subsubsection{Comparison with Direct Detection}

Figure~\ref{fig:DD} directly compares our net crust sensitivity to constraints from underground experiments searching for dark matter in nuclear recoils, for both spin-independent and spin-dependent scattering~\cite{Angloher:2015ewa,Ajaj:2019imk,Agnes:2018ves,Amole:2019fdf,Aprile:2019dbj}.
We also show the xenon neutrino floor \cite{Billard:2013qya,Ruppin:2014bra}, which denotes the cross section sensitivity below which neutrino backgrounds are expected to affect direct detection searches.
The shaded ``Full Crust'' band denotes the uncertainties in capture cross section arising from varying the neutron star mass, radius, rotation frequency $f_{\rm rot}$, and equation of state (EoS)\footnote{We did not consider neutron stars of mass $< 1.4 M_\odot$ as our equations of state predict for them a crust thickness $> 10\%$ the size of the star, invalidating the light-and-thin crust approximation that we assume throughout our work. 
Neglecting this stellar mass range only leads to conservative results, as a thicker crust implies greater optical depth and hence stronger sensitivities.}.
We find that the weakest sensitivity is provided by
\{$\Mstar = 2.3~M_\odot$, $\Rstar$ = 11.5 km, $f_{\rm rot}$ = 716 Hz, EoS = BsK21\}, and the strongest sensitivity by 
\{$\Mstar = 1.4~M_\odot$, $\Rstar$ = 11.75 km, $f_{\rm rot}$ = 0 Hz, EoS = BsK20\}. 

We have estimated the spin-dependent capture cross section shown in Fig.~\ref{fig:DD} by

\begin{enumerate}
 \item Scaling the spin-independent capture cross section by a factor of 
\begin{equation}
\left[\frac{4}{3}\frac{J_n+1}{J_n} (\langle S_n\rangle + \langle S_p\rangle)^2\right]^{-1}~,
\end{equation}
where $J_n = 1/2$, $\langle S_n\rangle = 1/2$ and $\langle S_p\rangle = 0$ for scattering on neutrons, and \\

 \item To calculate pasta scattering we have used the pasta structure factor in Eq.~\eqref{eq:Nscatterspasta} but have imposed $S_{\rm pasta} (q) \leq 1$, since we expect spin-dependent scattering to not be coherent over pasta nucleons, and
 
 \item To calculate quasi-elastic scattering on nucleons in the inner and outer crust, we follow the same procedure as spin-independent scattering, since in this case scattering occurs with individual nucleons.
\end{enumerate}

Although the PICO-60 limit applies to spin-dependent scattering on protons, we display it as it provides the current best sensitivity to spin-dependent dark matter interactions with nucleons at low dark matter masses. Furthermore, most spin-dependent models predict roughly equal rates of neutron and proton scattering.

For spin-independent scattering, we see that neutron crust heating is clearly more sensitive than direct detection for low dark matter masses ($\mdm \lesssim$ GeV); crustal capture easily proceeds for these masses (see Eq.~\eqref{eq:Eloss}), but at terrestrial experiments it is difficult to detect soft nuclear recoils above energy thresholds.
Direct detection experiments benefit from an $A^2$ enhancement in cross section from nuclear coherence, which helps them surpass neutron star crust sensitivity in some range of $\mdm$.
For spin-dependent scattering, we see that neutron crust heating is more sensitive than current direct detection searches for nearly all dark matter masses.
We note that these statements hold true for sub-MeV dark matter masses as well, where neutron star crust heating occurs through phonon excitations (see Fig.~\ref{fig:bounds1}).
For complementary bounds on dark matter-nucleon scattering at low dark matter masses, see Refs.~\cite{Erickcek:2007jv,
Gluscevic:2017ywp,
Xu:2018efh,
Cappiello:2018hsu,
Bringmann:2018cvk,
Bhoonah:2018gjb}.

For both spin-independent and dependent scattering, we have also shown the approximate optimistic reach of dark crustal heating. As old neutron stars are expected to be around 100 K hot, assuming future optimistic experimental sensitivity, the neutron star would have to be heated by dark matter to higher than about 100 K to have a differentiable signal. This optimistic reach is shown as the dotted ``100 K neutron star'' curve.

\subsubsection{Comparison with Outer Core Scattering}

Lastly, we remark on dark matter capture via scattering with the outer core of the neutron star.
This is the region marked by densities $0.5 - 2 \rho_0$, to distinguish it from the inner core with densities $> 2 \rho_0$, where exotic phases of matter are usually thought to appear.
It is not clear that exotic phases do not appear at outer core densities as well~\cite{Reid:1968sq,Shapiro:1983du,Witten:1984rs,Zacchi:2015lwa,Holdom:2017gdc}.
In addition, the neutrons (protons) in the outer core are expected to exist in a superfluid (superconducting) phase, so that dark matter could only scatter by exciting collective modes at low momentum transfers.

Therefore, while scattering on the outer core is subject to the same uncertainties as the inner core detailed in the introduction of this paper, nevertheless for completeness' sake we can estimate the capture cross section for the outer core with average density $\rho_{\rm OC}$ and thickness $d_{\rm OC}$ as
\begin{equation}
 10^{-44} \ {\rm{cm^{2}}} \left(\frac{1.5 \rho_{0}}{\rho_{\rm OC}}\right) \left(\frac{5 \ {\rm{km}}}{d_{\rm OC}}\right) {\rm{Min}}\left[1,\frac{p_{\rm Fermi}}{\mdm\gamma v_{\rm esc}}\right]~,
\end{equation} 
where the last term contains a Pauli-blocking factor for low momentum transfers, with $p_{\rm Fermi}$ the neutron Fermi momentum $\approx$~0.45~GeV~\cite{Baryakhtar:2017dbj}. Because this estimate will be affected by any novel nuclear dynamics in the outer core, for the rest of this paper we will focus on dynamics in the sub-nuclear crust. We leave a thorough exposition of the outer core, and associated uncertainties, to future work.

\section{Dark Matter Annihilation Heating}
\label{sec:ann}

After capture, dark matter can thermalize through repeated scattering with the crust and become confined to the volume underneath it. At this point, it will continue to oscillate in the stellar interior at semi-relativistic speeds. If the dark matter particles annihilate to Standard Model products, their mass energy is deposited into the star, heating it to temperatures even higher than that induced by kinetic heating. 
For instance, the temperature of our benchmark neutron star (Eq.~\eqref{eq:benchmarkstar}) due to dark matter kinetic-and-annihilation heating is 
\begin{equation}
T_\infty = 2470~{\rm K}~.
\end{equation}
This would have a significant effect on observational prospects at imminent infrared telescopes, in particular reducing the observation time by a factor of $\sim$10 and affecting the optimal choice of filters to be used. 
In this section we explore the possibility of dark matter annihilation post crustal capture in some detail, and show that hitherto-elusive scenarios such as supersymmetric Higgsino dark matter may be cornered thanks to this phenomenon.

The number of dark matter particles contained in the star evolves according to
\begin{equation}
 \frac{dN}{dt}=C_{\rm DM}-\frac{\sigvee N^{2}}{V}~,
\end{equation}
where $C_{\rm DM}$ is the capture rate given by $\dot{M}_{\rm DM}/\mdm$ from Eq.~\eqref{eq:masscapturerate}, 
$\sigvee$ is the thermally averaged dark matter annihilation cross section, and $V$ is the volume in which annihilation occurs, which we conservatively take to be the volume of the star. 
The solution to the above is given by
\begin{equation}
 N(t)=\sqrt{\frac{C_{\rm DM}V}{\sigvee}} 
 \tanh \left( \frac{t}{\tau_{\rm eq}}
      \right)~,
\end{equation}
where
\begin{equation}
 \tau_{\rm eq} = \sqrt{\frac{V}{C_{\rm DM}\sigvee}} 
 \label{eq:teq}
\end{equation}
is the characteristic timescale at which capture and annihilation equilibrate, after which $N(t)$ achieves a steady state.
For $t>\tau_{\rm eq}$ the  
net luminosity of the neutron star (from both kinetic and annihilation heating) is $(\dot{M}_{\rm DM} + \dot{E}_{\rm ann}) \propto C_{\rm DM} \mdm \gamma$.

$\sigvee$ can be expanded in terms of its partial wave contributions:
\begin{equation}
 \sigvee = (\sigma v)_0 \sum_{\ell = 0}^\infty a_\ell v_{\rm rel}^{2\ell}~,
 \label{eq:sigmavexpansion}
\end{equation}
where $v_{\rm rel}$ is the relative velocity of the annihilating dark matter particles, and $\ell = 0, 1, 2$ correspond respectively to the leading contributions from
$s$-wave, 
$p$-wave, 
$d$-wave modes.
For illustration we will consider scenarios where $\sigvee$ is dominated by a single mode such that the coefficients $a_\ell$~=~1 for that mode and $a_\ell$~=~0 for all other modes.
Note that the average relative velocity of dark matter particles oscillating within the star is approximately just the escape velocity, which is of order 0.1.
Using the approximate scaling $C_{\rm DM} \propto \Rstar^2$ from Eq.~\eqref{eq:masscapturerate} and $V \propto \Rstar^3$,
we see from Eq.~\eqref{eq:teq} that for annihilations dominated by the mode $\ell$, 
\begin{equation}
 \tau_{\rm eq} \propto \Rstar^{(3+\ell)/2}~.
 \label{eq:teqscaling}
\end{equation}

\begin{figure}[t]
\includegraphics[scale=0.7]{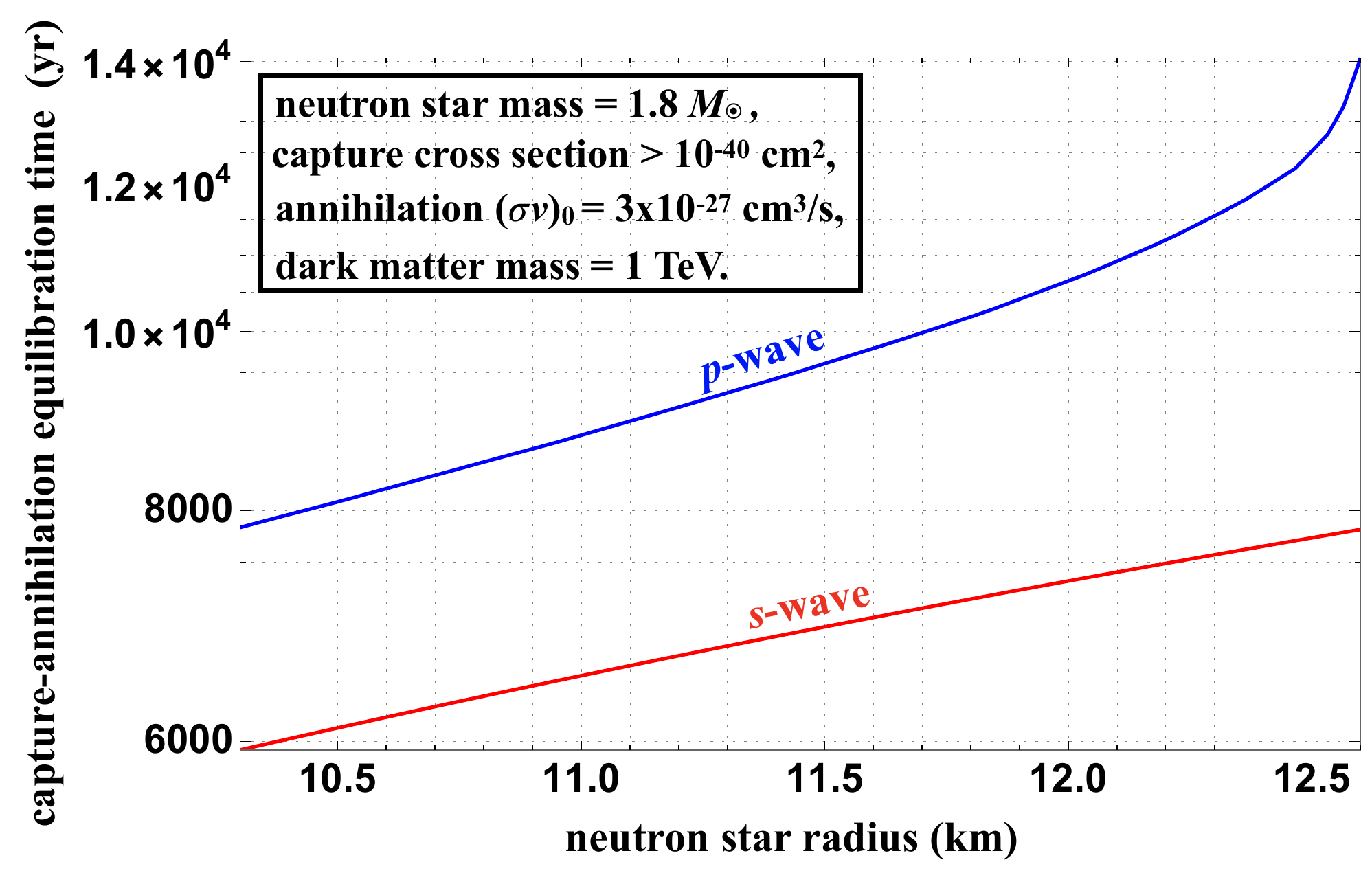} 
\caption{Capture-annihilation equilibrium timescale as a function of neutron star radius, for dark matter that fails to scatter and thermalize with the core of the star.
The timescale for both $s$-wave- and $p$-wave-dominated annihilations are shown.
Here we take the dark matter mass to be 1~TeV and a per-nucleon scattering cross section $> 10^{-40}$~cm$^2$, so that dark matter is guaranteed to capture in the crust (see Fig.~\ref{fig:bounds1}); the scaling cross section $(\sigma v)_0$ in Eq.~\eqref{eq:sigmavexpansion} is set to $3 \times 10^{-27}$~cm$^3$/s.
}
\label{fig:teq}
\end{figure}
 
Figure~\ref{fig:teq} shows $\tau_{eq}$ as a function of the radius of our benchmark 1.8~$M_\odot$ neutron star\footnote{We limit the plot to radii $\lesssim$~13~km since above that radius neutron stars are predicted to have masses $\lesssim M_{\odot}$, which are expected to have more massive and larger crusts, invalidating the light-and-thin crust approximation assumed throughout this work. We also remark that lighter neutron stars will be more efficient at capturing dark matter in their crusts since these have a greater optical depth.}, for $s$-wave and $p$-wave dark matter annihilations.
Here we fix $\mdm$~=~1~TeV and $(\sigma v)_0$~=~3$\times$10$^{-26}$~cm$^3$/s in Eq.~\eqref{eq:sigmavexpansion}, and assume a dark matter-nucleon cross section greater than $10^{-40}$~cm$^2$ so that all transiting dark matter is captured by the crust (see Fig.~\ref{fig:bounds1}).
As borne out by the scaling relation in Eq.~\eqref{eq:teqscaling}, the timescales shown would be shorter for smaller neutron stars, and for annihilations dominated by smaller partial wave modes.

We now note that were dark matter to scatter and even just partially thermalize with the core of the star, $\tau_{\rm eq}$ for $s$- and $p$-wave annihilations would be even smaller than that predicted by crust-only thermalization.
To see this, consider a partially thermalized ensemble of dark matter particles that is confined to a volume of radius $r_{\rm PT} < \Rstar$.
From the fact that the star's gravitational potential at this radius is $ (3G\Mstar/2\Rstar)(1-r^2_{\rm PT}/3\Rstar^2)$, the typical dark matter speed scales as $v_{\rm rel} \propto r_{\rm PT}$.
Using this and $V \propto r^3_{\rm PT}$ in Eq.~\eqref{eq:teq}, for dark matter annihilations dominated by the mode $\ell$ we have
\begin{equation}
\tau_{\rm eq} \propto r_{\rm PT}^{(3-2\ell)/2}~. 
\label{eq:teqscalingrth}
\end{equation}
Therefore for $\ell\leqslant 2$, the capture-annihilation equilibrium timescale scales positively with $r_{\rm PT}$: the smaller the thermalization volume, the quicker the equilibration. 
Conversely, for $\ell\geqslant 2$, this equilibration could take significantly longer for $d$-wave and higher-order modes.
We could also, of course, verify this statement in the limit of fully thermalized dark matter. 
In that case, for a core density $\rho_{\rm NS}$ dark matter would settle into a ``thermal volume" of radius $\ll$ the size of the star~\cite{Kouvaris:2010jy}:
\begin{equation}
 r_{\rm thermal} = \sqrt{\frac{9T}{4\pi G \rho_{\rm NS} \mdm }}\lesssim 10^{-2} \ {\rm{km}} \left(\frac{10^{15} \ {\rm{g/cm^{3}}}}{\rho_{\rm NS}}\right)^{1/2} \left(\frac{10^{3} \ \gev}{\mdm}\right)^{1/2} \left(\frac{T}{1000 \ {\rm{K}}}\right)^{1/2}~.
 \label{eq:rthermal}
\end{equation}
Since $v_{\rm rel} \propto \sqrt{T}$, the above equation implies $v_{\rm rel} \propto r_{\rm thermal}$ such that the scaling in Eq.~\eqref{eq:teqscalingrth} is once again true.

Some comments are in order regarding whether dark matter annihilation products efficiently heat the neutron star. 
So far we have assumed that this is the case. 
However, one might consider whether this holds if the annihilation products were neutrinos, whose mean free path in neutrons at saturation density $n_{\rm n}$ = 10$^{37}$~cm$^{-3}$ is~\cite{Raffelt:1996wa}
\begin{align}
L_\nu = (n_{\rm n} G_{\rm F}^2 E_\nu^2)^{-1} \simeq 10^{-1}~{\rm m}~\left(\frac{100~{\rm MeV}}{E_\nu}\right)^2,
\end{align}
where $G_{\rm F}$ is Fermi's constant and $E_{\nu}$ is the incident neutrino energy, equal to the dark matter mass for prompt $2\rightarrow 2$ annihilations. From this we see that we should expect dark matter annihilation to neutrinos, for the masses we have considered, to result in efficient neutron star heating.
Certainly for $\mdm \gtrsim$ 100 MeV these neutrinos do not escape the star, but rather scatter on the pasta layers of the inner crust and contribute to annihilation heating. 
In the future, it will also be interesting to consider dark matter annihihations to long-lived mediators~\cite{Batell:2009zp, Schuster:2009au, Schuster:2009fc, Meade:2009mu, Bell:2011sn, Feng:2015hja,Feng:2016ijc,Leane:2017vag,Albert:2018jwh,Bell:2017irk,Cermeno:2018qgu,Nisa:2019mpb,Niblaeus:2019gjk} or a second component of (nucleophilic) dark matter~\cite{Cui:2017ytb,Cherry:2015oca,McKeen:2018pbb}, which may also heat the crust rather than the core.

\subsection{Discovering Inelastic (Higgsino) Dark Matter}
\label{subsec:Higgsino}

Previously, we showed that dark matter capture by the neutron star crust alone is sufficient for capture-annihilation equilibrium within the lifetime of the star, for electroweak-sized annihilation cross sections. 
This means that the crust provides sensitivity to certain dark matter models that are difficult to probe at terrestrial experiments, such as dark matter that only scatters inelastically~\cite{TuckerSmith:2001hy}, and
dark matter that scatters through velocity-suppressed interactions~\cite{Dienes:2013xya}.
In this sub-section we illustrate this {\em crust} sensitivity by studying a particularly attractive model of inelastic dark matter, the supersymmetric Higgsino. We note that there is a recent complementary treatment of neutron star {\it core} capture of inelastic dark matter given in Ref.~\cite{Bell:2018pkk}.

For inelastic scattering on nucleons in the crust (we neglect coherent scattering on pasta for the moment), dark matter of mass $\mdm$ must scatter to an excited state of mass $\mdm + \delta$, with the inelastic cross section given by
\begin{equation}
 \sigma_{\rm n\chi} =  \sigma^0_{\rm n\chi} \sqrt{1-\frac{\delta}{\delta_{\rm max}}},
 \label{eq:inelasticxs}
 \end{equation}
where $\sigma^0_{\rm n\chi}$ is the (elastic) scattering cross section at vanishing mass splitting $\delta$, and $\delta_{\rm max}$ is the maximum mass splitting for inelastic scattering to proceed, given by 
\begin{equation}
\delta_{\rm max} = \frac{\mu_{\rm n\chi}v^2}
                        {2}~,
\label{eq:deltamax}
\end{equation}
with $\mu_{\rm n\chi}$ the neutron-dark matter reduced mass and $v \simeq v_{\rm esc}$.
For our benchmark star parameters, Eq.~\eqref{eq:benchmarkstar}, we have $\delta_{\rm max}$~=~200~MeV for $\mdm \gg m_{\rm n}$. 
Note that at terrestrial detectors $v \simeq 10^{-3}$, so that for a 100 GeV-heavy scattering target and a heavier dark matter projectile, $\delta_{\rm max} \simeq$~50~keV.
Thus neutron star crustal capture improves the ``inelastic frontier" of dark matter direct detection by $2-3$ orders of magnitude.

Higgsinos are the fermionic supersymmetric partners of the Standard Model Higgs field, 
 which undergo mass mixing with the superpartners of the weak gauge bosons (``binos" and ``winos") after electroweak symmetry breaking.
 If the bino and wino are decoupled from the Higgsino by setting their mass parameters (denoted by $M_1$ and $M_2$ respectively) much greater than the Higgsino mass parameter $\mu$, then a nearly pure Higgsino is a candidate for dark matter.
 Being an electroweak doublet, there are two neutral states at the mass scale $\mu$, denoted by $\chi_{1}$ (the dark matter state) and $\chi_{2}$ (an excited state), as well as a
charged Higgsino pair, $\chi^{\pm}$ (also potentially excited states).
For expressions of the mass splittings between $\chi_1$ and \{$\chi_{2},\chi^\pm$\} in terms of electroweak and supersymmetric parameters, see Ref.~\cite{Baryakhtar:2017dbj}.
For Higgsino mass $\mu$~=~1.1~TeV, Higgsino dark matter could be produced at the observed relic abundance, $\Omega_\chi h^2 = 0.11$, via thermal freeze-out from $s$-wave annihilations with $\sigvee \simeq 3 \times 10^{-27}$~cm$^2$~\cite{ArkaniHamed:2006mb}.

When kinematically allowed, the Higgsino $\chi_1$ can scatter inelastically to an excited state (by exchanging a $Z$ or $W^\pm$ boson) with an electroweak-sized cross section $\sigma^0_{\rm n \chi} \simeq 10^{-39}$~cm$^2$ in Eq.~\eqref{eq:inelasticxs}.
From Fig.~\ref{fig:bounds1}, this implies that thermal Higgsino dark matter of mass 1.1 TeV will:

\begin{enumerate}
 \item be captured by the outer crust whose capture cross section $\simeq$~10$^{-40}$~cm$^2$; as mentioned in Sec.~\ref{subsec:captureremarks}, as the outer crust is the most empirically understood layer, this is a robust prediction for electroweak dark matter such as the thermal Higgsino,
\item further scatter with and deposit energy in the inner crust (including the pasta layer), whose capture cross section $\simeq$~10$^{-41}$~cm$^2$.
\end{enumerate}

The above stages of capture would occur for most values of the mass splitting $\delta < \delta_{\rm max}$ in Eq.~\eqref{eq:inelasticxs}, and any deviations would only occur for $\delta$ within 1\% of $\delta_{\rm max}$.
Therefore, most of the thermal Higgsino parameter space corresponds to the curve in Fig.~\ref{fig:teq} denoting $s$-wave annihilations for $\mdm = 1$~TeV, {\em i.e.} crustal capture implies that capture-annihilation equilibrium is expected in timescales of less than $10^4$~years.
This in turn implies that thermal Higgsino dark matter with inelastic mass splitting less than 200 MeV is practically guaranteed to give rise to both kinetic and annihilation heating of neutron stars, significantly increasing the prospects of its discovery. \textit{This is one of the main results of our paper.}\\

As detailed in the Introduction, this result provides a key motivation for treating neutron star crusts as dark matter captors.
Even if the stellar core were not exotic, dark matter scenarios like the supersymmetric Higgsino may never thermalize with the core through repeated scattering.
This can happen for a number of reasons:

\begin{enumerate}
 \item As the Higgsino loses kinetic energy, $\delta_{\rm max}$ decreases (see Eq.~\eqref{eq:deltamax}), so that inelastic scattering eventually becomes forbidden; 
 \item In addition to inelastic scattering, Higgsino dark matter can also scatter {\em elastically} ($\chi_1 n \rightarrow \chi_1 n$) 
via a higher-order amplitude, however the cross section for this process is currently unknown, and only an upper bound has been computed~\cite{Hill:2014yka,Hisano:2011cs}: 
$\sigma(\chi_1 n \rightarrow \chi_1 n) \lesssim 10^{-48}$~cm$^2$ for $\mdm$~=~1.1~TeV.
\end{enumerate}

As no lower bound exists on this cross section, the Higgsino might in principle never thermalize with the core within the present age of the universe.
As a result, Higgsino dark matter may annihilate in a volume much larger than the thermal volume given by Eq.~\eqref{eq:rthermal}, {\em i.e.} much less efficiently. However, we have established that \textit{Higgsinos will annihilate anyway}: even in the limit of capture in the crust and no thermalization with the core, capture-annihilation equilibrium is attained very quickly as shown in Fig.~\ref{fig:teq}, so that annihilation proceeds at its maximal rate.

\section{Summary and Conclusions}
\label{sec:conc}

In this work, we have investigated the different scattering processes via which particle dark matter would capture in neutron star crusts.
Our results for each crustal layer can be summarized as follows:

\begin{itemize}
 \item We have calculated dark matter scattering and capture by the outer crust of neutron stars. 
We find that this is dominated by dark matter-nucleon scattering, with cross section sensitivity of about $10^{-40}$ cm$^2$, for dark matter masses ranging from 10 MeV to 1 PeV. 
The physics of the outer crust is best-understood of all the layers, which is promising as electroweak-scale dark matter is expected to be within the sensitivity of this layer, with scattering cross sections of around $10^{-40}$ cm$^2$.

 \item For the first time, we have calculated dark matter scattering with the inner crust of neutron stars. 
 In the outer parts of the inner crust, there is no pasta phase, and this region is dominated by dark matter-nucleon scattering, with cross section sensitivity of about $10^{-41}$ cm$^2$ for dark matter masses ranging from 10 MeV to 1 PeV.
 
For dark matter masses 10 eV~$\lesssim \mdm \lesssim$~1~MeV, we have also considered dark matter scattering with the neutron superfluid, which leads to phonon excitations. This provides cross section sensitivity of $10^{-39} - 10^{-34}$~cm$^2$.
 
  \item For the first time, we have calculated dark matter scattering with nuclear pasta. 
 Nuclear pasta phases are predicted to appear in the innermost region of the inner crust.  
Dark matter-pasta interactions provide the best nucleon cross section sensitivity in the whole crust, ranging from $10^{-43}$ cm$^2$ to  $10^{-41}$ cm$^2$ for 10 MeV -- 1 PeV dark matter masses. 
The strongest sensitivity is obtained near 100 MeV, where there is substantial enhancement in cross sections due to coherent dark matter-pasta scattering. 
Below this mass, there is decoherence leading to weaker sensitivities. 
Above this mass, individual nucleons become resolved, resulting in quasi-elastic scattering on free nucleons.
 
 \item Throughout all the layers, we find that dark matter-nucleus scattering is suppressed because of the combined effect of the Helm form factor, which suppresses interactions at high momentum transfers, and the static structure factor of the lattice, which suppresses interactions at low momentum transfers. 
Instead, dark matter is able to scatter directly on nucleons, both in nuclei and in the fluid phase, as long as the energy transfer to nucleons is larger than the respective binding energy per nucleon or the superfluid energy gap.

\end{itemize}

Our results also have important implications for dark matter thermalization in neutron stars. If dark matter-core interactions do not exist or are suppressed, once dark matter particles have settled in orbits contained below the crust, no further energy would be lost. 
This would limit the amount of energy imparted by the dark matter particles to about a fourth of their mass. 
However, if the dark matter particles, which remain oscillating inside the star, annihilate to Standard Model particles, these annihilation products transfer the remaining energy to the star. 
We find that with crust-only interactions, the time required to reach capture-annihilation equilibrium is much shorter than the ages of old neutron stars for thermal annihilation cross sections.  
We have therefore shown that dark matter can thermalize in neutron stars, without having to consider the largely uncertain physics of the core.

We have found that substantial sensitivity is achievable using the crust for inelastic dark matter models. 
For direct detection experiments, where dark matter travels at non-relativistic speeds, the maximum mass splitting sensitivity is $\delta \sim$~100~keV.
For neutron stars crusts, we find that the maximum mass splitting for electroweak scale dark matter to be captured and thermalized, and subsequently annihilated inside, is $\delta \gtrsim 200 \ \mev$. 
The large difference in these sensitivities highlights the importance of considering dark matter-crust interactions. 
We showed as an explicit example that this sensitivity is achievable in a thermal Higgsino dark matter model.

We have emphasized in this work that constraints arising from dark matter-crust interactions are more robust than those obtained in a typical treatment using core-only scatters. 
Although we find  that capture cross sections are larger for the crust alone compared to the core~\cite{Baryakhtar:2017dbj}, we have presented results that are not prone to the same uncertainties present at super-nuclear densities. 
This provides a far-reaching objective for forthcoming infrared telescopes: they are {\em guaranteed} to observe or constrain dark matter heating of neutron stars via crust-only scatters, which exist even if neutron star cores turn out to be exotic or suppress dark matter interactions.

The dynamics of scattering on different layers of a star stabilized by quantum degeneracy pressure may have broad applications, for instance in stars made chiefly of dark matter that may capture Standard Model particles~\cite{Curtin:2019lhm,Curtin:2019ngc}. 
We \textit{leaven} this and other applications of dark matter scattering on compact star crusts to future exploration.

\section*{Acknowledgments}

We are grateful to 
David Curtin,
Graciela Gelmini,
Aniket Joglekar,
Charles Horowitz,
David Morrissey,
Rana Nandi,
Sanjay Reddy,
and Michael Wagman
for discussions.
We thank the anonymous referee for their careful review of our work.
The work of RKL is supported by the Office of
High Energy Physics of the U.S. Department of Energy
under Grant No. de-sc00012567 and de-sc0013999. 
The work of JA, JB and NR is supported by the Natural
Sciences and Engineering Research Council of Canada (\acro{NSERC}).
T\acro{RIUMF} receives federal funding
via a contribution agreement with the National Research Council Canada.
This work was also prepared in part at the Aspen Center for Physics, which is supported by National Science Foundation grant PHY-1607611.

\appendix

\section{Detailed Structure of Neutron Star Crusts}
\label{app:crust}

In this appendix we elaborate on the structure of neutron star crusts briefly described in Sec.~\ref{sec:struc}.
Our discussion here also illuminates various aspects of dark matter capture in different layers of neutron star crusts as described in Sec.~\ref{sec:capture}.

\subsection{Overall Structure}

We begin by reviewing the overall structure of a neutron star.
This structure is defined by the equation of hydrostatic equilibrium (Tolman-Oppenheimer-Volkoff equation) for the star,
\begin{equation}
\frac{dP}{dr}=-\frac{G\rho{m}}{r^{2}}\left(1+\frac{P}{\rho}\right)\left(1+\frac{4\pi{P}r^{3}}{m}\right)\left(1-\frac{2Gm}{r}\right)^{-1}~, 
\label{eq:tov}
\end{equation}
and the conservation of mass,
\begin{equation}
 \frac{dm}{dr}=4 \pi r^{2} \rho~,
\end{equation}
where $r$ is the radial coordinate, 
$P(r)$ is the pressure and
$\rho(r)$ is the gravitational mass density at $r$,
and $m(r)$ is the gravitational mass enclosed in a sphere of radius $r$. 
The stellar surface at radius $\Rstar$ is defined by the condition $P(\Rstar)=0$, which in turn gives the total gravitational mass of the star $\Mstar=4\pi{\int_{0}^{\Rstar}{\rho(r) r^{2} dr}}$.

To obtain a closed system of equations, an equation of state is required to relate pressure to density. 
In this work, we use an equation of state describing the neutron star crust with a   nuclear interaction Hamiltonian; in particular, we use analytical representations of $P(\rho)$ known as the Brussels-Montreal equations of state~\cite{Goriely:2010bm,Chamel:2010rw,Potekhin:2013qqa}.
Among its versions, we have used ``BsK21" as our standard equation of state -- some other versions are disfavored by observations~\cite{Radice:2018ozg,Pearson:2018tkr}. Later, we also employ a few other ``BsK'' models to show that our results are robust when this choice is varied.
We use this equation of state to solve Eq.~\eqref{eq:tov}, which we in turn use  to compute the optical depth of each crustal layer in Eq.~\eqref{eq:scatt-nmbr}.

Next, we inspect in detail the structure of the outer and inner crusts, including the pasta phase.

\subsection{Outer Crust}
\label{subsec:outer}

The outer crust is composed of nuclei forming a body-centered-cubic Coulomb crystal, interspersed with a degenerate and nearly-free relativistic gas electrons. 
The electron degeneracy contributes dominantly to the pressure, while nuclei contribute dominantly to the density. 
Deep in the crust, nuclei become increasingly neutron-rich due to inverse $\beta$ decay, $e^- + p \rightarrow n + \nu_e$ due to increased electron chemical potential.
The outer crust terminates when the density and pressure become so high that free neutron states begin to appear. 

Nuclear masses have been measured at densities far below this ``neutron drip" point,  and are used as inputs to accurately determine the composition of the outer crust~\cite{1971ApJ...170..299B}. 
However, closer to the neutron drip line, nuclear masses are unknown so that semi-empirical formulae and extrapolations must be used~\cite{Ruester:2005fm}.
For the purposes of this work, these uncertainties are unimportant as predictions for the mass number of each nuclear species differ only by a few units. 
Therefore, using the results of Refs.~\cite{1971ApJ...170..299B,Ruester:2005fm,Sharma:2015bna,Potekhin:2013qqa,Haensel:1993zw,Wolf:2013ge,Douchin:2001sv,RocaMaza:2008ja,Pearson:2011zz}, we model the outer crust as three layers of approximately constant composition\footnote{There are small discontinuities in density as we transition between elemental species in these layers, however these discontinuities would smoothen out in a lattice formed by an admixture of elements~\cite{1982ApJ...253..839J}.} as summarized in Table~\ref{tab:outercrust}. Note that the elements shown are just what is common across most equation of states, and other elements can be present depending on the equation of state choice.

\begin{table}[]
\centering
\begin{tabular}{|c|c|c|c|c|}
\hline
layer & elements & mass number $A$ & density ($\rm{g/cm^{3}}$) & electron chemical potential ($\mev$) \\ \hline
1 & Fe, Ni & 56 & [$10^{6}$, $2 \times 10^{9}$] & [$0.5$, $5$] \\ \hline
2 & Ni, Kr, Zn, Ge & 82 & [$2 \times 10^{9}$, $7 \times10^{10}$] & [5, 17] \\ \hline
3 & Mo, Zr, Kr, Se & 118 & [$7 \times 10^{10}$, $3 \times 10^{11}$] & [17, 26] \\ \hline
\end{tabular}
\caption{Simplified model of the layers of the outer crust of a non-accreting neutron star used in this work.
See Sec.~\ref{subsec:outer} for more details.}
\label{tab:outercrust}
\end{table}

\begin{figure}[t] 
\includegraphics[scale=0.7]{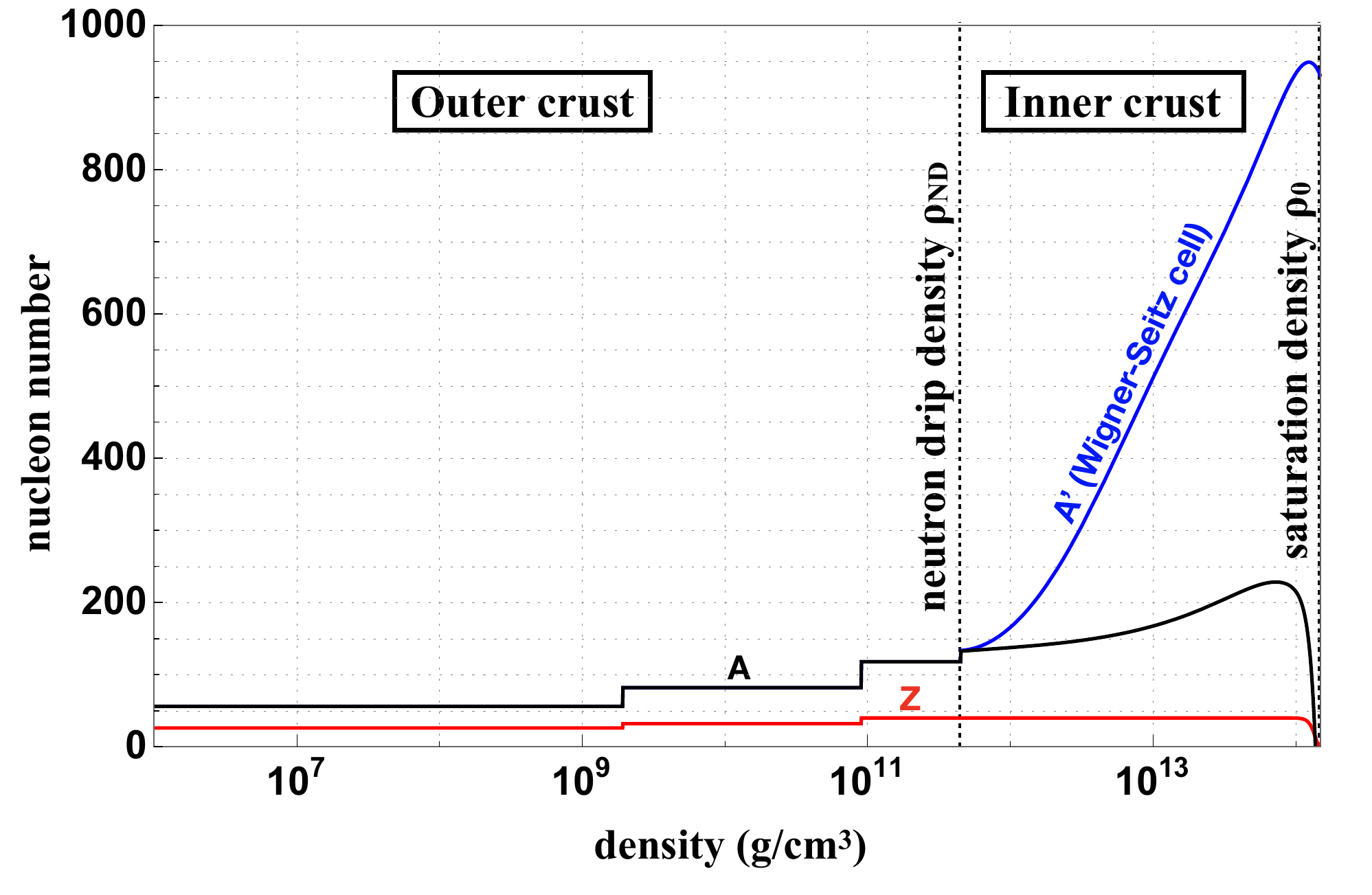} 
\caption{Nucleon number as a function of neutron star crust density.
See text for details.
}
\label{fig:za}
\end{figure}

\subsection{Inner Crust}
\label{subsec:inner}

The transition to the inner crust is marked by the neutron drip line, beyond which a fraction of neutrons become unbound from nuclei. 
Up to densities $\sim 0.1$ times the saturation density $\rho_0$, the inner crust comprises of heavy, neutron-rich nuclei (also known as proton clusters) forming a lattice, along with both an electron gas and a dripped-neutron gas. 
Such a system is inaccessible to terrestrial experiments, hence the composition of the inner crust is far more uncertain than its outer counterpart. 
Studies of this region are therefore limited to theoretical calculations only. 
These include classical models such as the Compressible Liquid Drop Model~\cite{Ravenhall:1972zz,Pethick:1995di,Douchin:2000kx}, 
semi-classical calculations such as the Thomas-Fermi approximation~\cite{Oyamatsu:1993zz}, and many-body quantum calculations~\cite{NEGELE1973298,SIEMENS1971561}.

As a neutron star cools down over time, the dripped neutrons are expected to form a superfluid phase~\cite{MIGDAL1959655,baym1969superfluidity,Pethick:2010zf}. 
These superfluid neutrons could significantly affect various properties of the neutron star. 
They could also cause pulsar glitches, which are considered as observational evidence of the existence of this superfluid phase~\cite{baym1969superfluidity,pines1985superfluidity}. 

In the inner crust, one could distinguish between the nucleons bound in proton clusters and the total population of nucleons including the dripped neutrons. 
The mass number $A'$ of nucleons enclosed in a Wigner-Seitz cell of the lattice is typically $\gg$ the mass number $A$ of nuclei in the inner crust.

\underline{\bf \em Pasta} 
At densities close to $\rho_0$, nuclei under extreme pressure turn inside out, marking the onset of homogeneous nuclear matter~\cite{BAYM1971225,LATTIMER1985646}.
This has led to the prediction of the so-called nuclear pasta phase at the bottom of the inner crust~\cite{Ravenhall:1983uh,10.1143/PTP.71.320,10.1143/PTP.72.373,Williams:1985prf,Lorenz:1992zz,Oyamatsu:1993zz}. 
This comprises non-spherical phases of nuclear matter such as thin tubes, slabs or spherical bubbles.
Nuclear pasta are confined to a thin layer, yet they constitute a significant fraction of the crustal mass as they span densities of $0.1 - 1 \ \rho_{0}$. 
They may also impact several properties of the neutron star such as its thermal and electrical conductivity, and the elasticity and neutrino opacity of the crust, the latter of which is relevant for the physics of core-collapse supernovae and neutron star cooling~\cite{Horowitz:2005zb,Sonoda:2007ni,ALCAIN2017183,Roggero:2017pag,Nandi:2017aqq}.
A number of computational studies of pasta phases have been recently carried out for different temperatures, densities and proton fractions, which the interested reader may find in Refs.~\cite{Magierski:2001ud,Gogelein:2007pb,Sonoda:2007sx,Iida:2001xy,Watanabe:2006qf,Schneider:2013dwa}.  

The inner crust terminates when the density reaches $\rho_{0}$, beyond which nuclei ``melt" into uniform nuclear matter and possibly other exotic phases of matter, which form the core of the neutron star.

Figure~\ref{fig:za} shows $A'$, $A$ and the proton number $Z$ of nuclei as a function of densities spanning the entire crust, using the equation of state BsK21.

\section{Discerning Crust vs. Core Dark Heating}
\label{app:discern}

\begin{figure}[t] 
\includegraphics[scale=0.6]{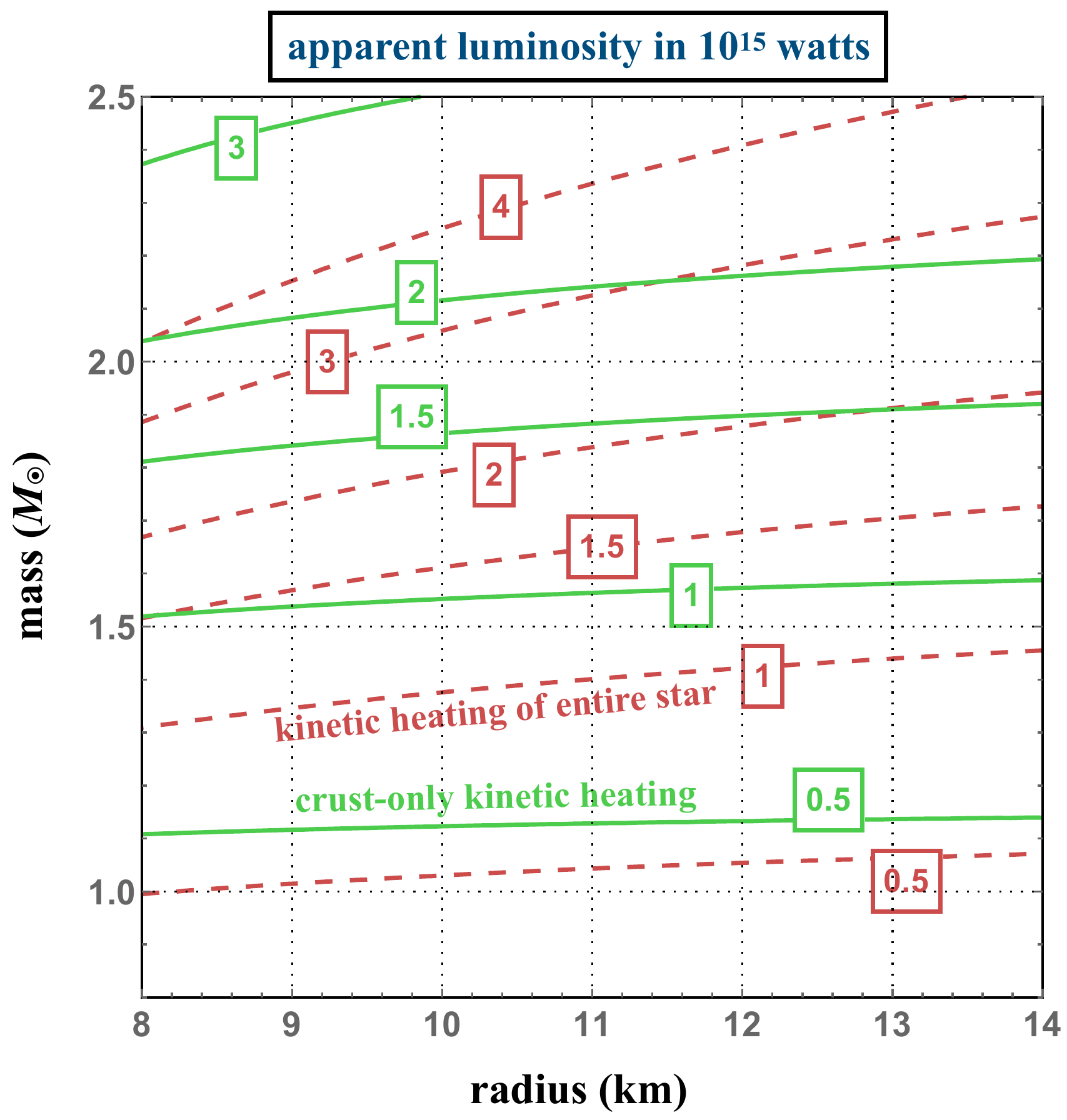} 
\caption{Apparent luminosities of neutron stars as a function of stellar mass and radius, for dark matter kinetic heating of only the crust and for the entire star.}
\label{fig:discern}
\end{figure}

Figure~\ref{fig:discern} shows contours of the apparent luminosity (in petawatts) of neutron stars equal to $4\pi\Rstar^2 T^4_\infty/\gamma^2$, plotted for various stellar masses and radii, for both crust-only and entire-star dark kinetic heating. This shows that, if it is possible to obtain sufficiently sensitive observations using infrared telescopes, neutron star crust-only heating of dark matter could potentially be discerned from heating of the entire star (such as in Eq.~\eqref{eq:T_signal}).
Both heating mechanisms are functions of $\Mstar$ and $\Rstar$. 
However, sensitivity to this scenario may be hampered by large uncertainties associated with such a measurement, from factors such as the equation of state, telescope systematics and statistical errors. We leave a detailed study to future work. For minimality, we have not shown regions of parameter space disfavored by causality considerations and data from fast-rotating neutron stars; see Ref.~\cite{Baryakhtar:2017dbj}.

\section{Dark Matter Thermalization with the Neutron Star Crust}
\label{app:thermstage1}

In Sec.~\ref{sec:capture} we discussed the conditions for dark matter capture. 
Once these conditions are met, dark matter particles scatter with the crust repeatedly, losing energy until they sink beneath the crust.
In this appendix we estimate the timescale over which this thermalization process occurs.

First note that the average energy lost by dark matter for each transit through the crust is
\begin{equation}
\Delta{E}_{\rm ave} \simeq \tau_{\rm DM} \times \Delta{E}~,
\label{eq:Elossave}
\end{equation}
 where $\tau_{\rm DM}$ is the optical depth given by Eq.~\eqref{eq:scatt-nmbr} and $\Delta{E}$ is the energy lost per scatter, given by Eq.~\eqref{eq:Eloss}.
Next, consider the radial trajectory of a dark matter particle oscillating across the neutron star.
This trajectory is equal to an ellipse with eccentricity 1 and semi-major axis $r$/2, where $r$ is the ``apastron", the maximum distance of the particle from the centre of the star, given in terms of the particle energy by
\begin{equation}
 r(E) = \frac{2GM}{1-(E/\mdm)^{2}}~.
 \label{eq:aphelion}
\end{equation}
The orbital period $t_{\rm orb}$ for this particle is then the same as that of a circular orbit of radius $r$/2, hence from Kepler's Third Law we have
\begin{equation}
 t_{\rm orb}= \eta^{-1} \frac{\pi r^{3/2}}{\sqrt{2G\Mstar}}~,
 \label{eq:torb}
\end{equation}
where $\eta = \sqrt{2}$ accounts for the particle trajectory through a homogeneous sphere~\cite{misner1973gravitation}. 
From Eqs.~\eqref{eq:Elossave} and \eqref{eq:torb} the rate of energy loss is
\begin{equation}
 \frac{dE}{dt} \approx \frac{2\sqrt{G\Mstar}}{\pi r(E)^{3/2}} \ \tau_{\rm DM} \Delta{E}~.
 \label{eq:elossrate}
\end{equation}

The total (proper) time for thermalization can now be obtained by using Eq.~\eqref{eq:aphelion} to integrate Eq.~\eqref{eq:elossrate} from an initial energy $\mdm + \mdm v^2_{\rm halo}/2$ to a final energy $\mdm \gamma^{-1}$, which is the gravitational binding energy at the star surface.
Performing this integration for $\Mstar = 1.8M_{\odot}$ and $\Rstar = 12.5 \ \rm{km}$, we have for $\mdm \gg m_{\rm n}$ the thermalization timescale:
\begin{equation}
 \tau_{\rm therm} = 5 \times 10^{4} \ {\rm{s}} \left(\frac{10^{-42} \ \rm{cm^{2}}}{\sigma_{\rm n\chi}}\right) \left(\frac{\mdm}{10^6~{\rm GeV}}\right)~,
\end{equation}
which is normalized to the largest $\mdm$ we consider and the net capture cross section of the crust for $\mdm \gtrsim$~1~GeV.
This is a very short timescale, and could be shorter for $\mdm = \mathcal{O}$(100 MeV) due to the coherent response of the pasta phases.
(As for $\mdm \ll m_{\rm n}$, it can be seen from Eq.~\eqref{eq:Eloss} that a single scatter deposits most of the dark matter kinetic energy into the crust.)

As discussed in Sec.~\ref{sec:capture} the total energy deposited by a dark matter particle through the thermalization process $\simeq \mdm (1-\gamma^{-1})$. 
An additional, small amount of energy will be deposited as the dark matter orbit shrinks further through the crust. 
Using the fact that the crust thickness $d_{\rm crust} \ll \Rstar$, and
taking the difference in the binding energies at the top and bottom of the crust, we find:
\begin{eqnarray}
\nonumber \Delta{E}_{\rm crust} &=& \mdm \left(\sqrt{1-\frac{2G\Mstar}{\Rstar}}-\sqrt{1-\frac{2G\Mstar}{\Rstar-d_{\rm crust}}}\right) \\
  &\simeq& \mdm \frac{(1-\gamma^{-2})}{2\gamma} \frac{d_{\rm crust}}{\Rstar} + \mathcal{O}\left(\left(\frac{d_{\rm crust}}{\Rstar}\right)^2\right)~,
 \label{eq:Ecrust}
\end{eqnarray}
which is a small fraction of the dark matter kinetic energy at the star surface, $\mdm (\gamma -1)$.
Here we have used the expression for the binding energy at the star surface for the interior as well; even though the metric in this region deviates from the typical Schwarzschild metric for a concentrated mass, this is a good approximation since the crust is very thin compared to the star.

\newpage
\bibliographystyle{JHEP}
\bibliography{cdm}

\end{document}